\documentclass[printer]{aa} 

\usepackage{graphicx}
\usepackage{txfonts}
\begin{document}

   \title{Future capabilities of CME polarimetric 3D reconstructions with the METIS instrument: A numerical test.}


\author{P. Pagano\inst{1} \and A. Bemporad\inst{2} \and D.H. Mackay\inst{1}}
\institute{{School of Mathematics and Statistics, University of St Andrews, North Haugh, 
St Andrews, Fife, Scotland KY16 9SS, UK.}
\and
{INAF-Osservatorio Astrofisico di Torino, via Osservatorio 20, 10025 Pino Torinese (TO)}
}

   \date{}

 
  \abstract
   {Understanding the 3D structure of coronal mass ejections (CMEs) is crucial for understanding the nature and origin of solar eruptions.
However, owing to the optical thinness of the solar corona we can only observe the line of sight integrated emission.
As a consequence the resulting projection effects hide the true 3D structure of CMEs.
To derive information on the 3D structure of CMEs from white-light (total and polarized brightness) images, the polarization ratio technique is widely used.
The soon-to-be-launched METIS coronagraph on board Solar Orbiter will use this technique to produce new polarimetric images.}
   {This work considers the application of the polarization ratio technique to synthetic CME observations from METIS.
In particular we determine the accuracy at which the position of the centre of mass, direction and speed of propagation, and  the column density
of the CME can be determined along the line of sight.}
   {We perform a 3D MHD simulation of a flux rope ejection where a CME is produced.
From the simulation we (i) synthesize the corresponding METIS white-light (total and polarized brightness) images and (ii)
apply the polarization ratio technique to these synthesized images and compare the results with the known
density distribution from the MHD simulation.
In addition,  we use recent results that consider
how the position of a single blob of plasma is measured depending on its projected position in the plane of the sky.
From this we can interpret the results of the polarization ratio technique
and give an estimation of the error associated with derived parameters.}
   {We find that the polarization ratio technique reproduces with high accuracy the position of the centre of mass along the line of sight.
However, some errors are inherently associated with this determination.
The polarization ratio technique also allows information to be derived 
on the real 3D direction of propagation of the CME.
The determination of this is of fundamental importance for future space weather forecasting.
In addition, we find that the column density derived from white-light images is accurate and 
we propose an improved technique where the combined use of
the polarization ratio technique and white-light images
minimizes the error in the estimation of column densities.
Moreover, by applying the comparison to a set of snapshots of the simulation
we can also assess the errors related to the trajectory and the expansion of the CME.}
   {Our method allows us to  thoroughly test the performance of the polarization ratio technique and
allows a determination of the errors associated with it, which means that  it can be used to
quantify the results from the analysis of the forthcoming
METIS observations in white light (total and polarized brightness).
Finally, we describe a satellite observing configuration relative to the Earth
that can allow the technique to be efficiently used for space weather predictions.}   

   \maketitle
%


\section{Introduction}

Coronal mass ejections (CMEs) are violent ejections of plasma and magnetic flux from the solar corona.
They exhibit a wide diversity of shapes and properties
and there are many unanswered questions on their origin and impact on space weather.
White-light observations from coronagraphs
are the most common way to detect and study CMEs.
However, their
3D structure is not fully understood because the solar corona is optically thin.
As a consequence we observe the 2D projection of CMEs onto the plane of the sky.
Many previous studies have tried to overcome this difficulty
by inferring either the density distribution along the line of sight (LOS) or
the location from where the CME emission originates.
\citet{Thernisien2011} present a complete survey
of the most commonly applied techniques, which include
forward modelling \citep{Thernisien2009},
polarimetric \citep{MoranDavila2004},
spectroscopic \citep{Raymond2002},
and direct inversion \citep{Frazin2009} techniques.
More recently, unique information on the 3D structure of CMEs have been derived thanks to the observations acquired by the twin STEREO spacecraft.
In this paper we focus on the validation of the polarimetric technique \citep{MoranDavila2004}.
This technique takes advantage of the fact that
the total and polarized brightness in white light from Thomson scattering has
a different dependency on the angle between the incoming scattered beam and the observer.
\citet{Moran2010} have already compared the results of the polarimetric technique with
the triangulation technique using STEREO spacecraft observations.
In addition, \citet{Mierla2011} performed an analysis of the limitations of the polarization technique from an observational point of view.
More recently, \citet{Dai2014} analysed in detail the ambiguities that arise in the application of this technique and suggested
a method for correctly reproducing the CME morphology.

The coronagraphic images and the stereoscopic observations taken from space have proven
to be very useful in the reconstruction of the 3D propagation directions of CMEs.
The determination of this information, as well as the predicted arrival time at 1 AU, is very important.
It is required in order to forecast the possible impact of interplanetary CMEs with the Earth's magnetosphere,
thus the possible occurrence of geomagnetic storms.
Before the availability of STEREO data,
many different methods for inferring the CME propagation direction were developed.
Examples are the cone model technique applied to LASCO data
\citep{Zhao2002, Michalek2003}
and the polarization ratio technique \citep{MoranDavila2004}.
Since the launch of STEREO many other methods have been proposed
such as forward modelling \citep{Thernisien2006},
analysis of height-time diagrams (Mierla et al. 2008), and more recently the ellipsoid model \citep{Schreiner2013}.
\citet{Zuccarello2012b} have used the triangulation technique to study the deflection of a CME
during its path in the solar corona.

In 2018 the new satellite Solar Orbiter,
with the Multi Element Telescope for Imaging and Spectroscopy (METIS) coronagraph \citep{Antonucci2012,Fineschi2013} will
be able to provide a new view of CMEs.
The Solar Orbiter mission makes it is very important to understand the validity
and the possible difficulties of the polarization
ratio technique for the derivation of 3D CME structure and propagation direction.
Specifically, METIS will observe the coronal emission in polarized
white light (bandpass between 580 - 640 nm) and in the UV (H I Lyman-alpha 1216 $\AA$)  at the same time.
As Solar Orbiter orbit will not be coplanar with the Earth orbit,
it will observe CMEs from a variety of angles
and from different heliocentric distances, approaching the Sun to 0.29 AU.
As a result the polarization ratio technique will be widely used to reconstruct 
the 3D morphology of observed CMEs.

The goal of this work is to apply the polarization ratio technique
to a well-known distribution of density produced from a MHD simulation of 
the ejection of a magnetic flux rope.
The ejection of magnetic flux ropes 
is believed to be one of the main progenitors of CMEs
and this scenario has been confirmed by \citet{Pagano2013a} and \citet{Pagano2013b}.
We first synthesize the white-light total and polarized brightness images
of the resulting CME as would be seen in the METIS field of view when the spacecraft is at its closest approach to the Sun.
Next we apply the polarization ratio technique to these images
and we compare the results of the polarization ratio technique 
with the original plasma distribution given by the MHD simulation.
This will assess the accuracy and the limitations of the technique
in the 3D reconstruction and trajectory determination of CMEs.
Finally, we repeat the same procedure for a set of snapshots of the MHD simulation to 
assess the accuracy of the technique in determining the CME properties describing its time evolution,
such as trajectory, expansion, and velocity.

This paper is the second study addressing this topic:
In \citet{BemporadPagano2015} we describe how
the position of a single blob of plasma is measured using the polarization technique
and estimate the error associated with the measurement.
The present paper is a continuation of that study where we apply the same
processes to a more realistic scenario using MHD simulations of
a flux rope ejection.
The aim is to show how the polarization ratio technique and the features of 
\citet{BemporadPagano2015} can be used to interpret polarization ratio images of CMEs.

The structure of the paper is as follows:
In Sect.\ref{mhdsimulation} we describe the MHD simulation;
in Sect.\ref{metisobservation} we illustrate how METIS would observe the CME produced in the simulation in white light;
in Sect.\ref{result} we analyse the results of the polarization ratio technique applied to the MHD simulations;
in Sect.\ref{discussion} we discuss our results and draw conclusions.

\section{MHD Simulation}
\label{mhdsimulation}

To simulate the occurrence of a CME
and produce a plasma density distribution
that can assess the validity of the polarization ratio technique
we perform a 3D MHD simulation of a flux rope ejection.
The plasma density distribution from the resulting ejection is
used to synthesize METIS observations in white light.
The simulation is similar to the ones already presented in 
\citet{Pagano2013a}, \citet{Pagano2013b}, and \citet{Pagano2014}.
In these studies a flux rope is first formed through the simulation technique
of \citet{MackayVanBallegooijen2006A},
which is used as an initial condition of the MHD simulation.
In particular, we use the magnetic configuration of Day 19
of the simulation
of \citet{MackayVanBallegooijen2006A}
when a flux rope has formed and is about to erupt.
Section 2.2 of \citet{Pagano2013a} describes in detail how the magnetic field distribution is imported from the
Global Non-Linear Force-Free Field (GNLFFF)model of \citet{MackayVanBallegooijen2006A} to the MHD simulations.
\citet{Pagano2013b} identified the parameter space where the flux rope is ejected.
Finally,  \citet{Pagano2014}  shows how a realistic density and temperature 
distribution can be constructed to account for the presence of a magnetic flux rope.
In the present work, the simulation parameters are chosen to generate
a CME in the MHD simulation.

We use the MPI-AMRVAC software~\citep{Porth2014}, to solve the MHD equations
where external gravity is included as a source term,
\begin{equation}
\label{mass}
\frac{\partial\rho}{\partial t}+\vec{\nabla}\cdot(\rho\vec{v})=0,
\end{equation}
\begin{equation}
\label{momentum}
\frac{\partial\rho\vec{v}}{\partial t}+\vec{\nabla}\cdot(\rho\vec{v}\vec{v})
   +\nabla p-\frac{(\vec{\nabla}\times\vec{B})\times\vec{B})}{4\pi}=\rho\vec{g},
\end{equation}
\begin{equation}
\label{induction2}
\frac{\partial\vec{A}}{\partial t}=\vec{v}\times\vec{B}
\end{equation}
\begin{equation}
\label{induction3}
\vec{B}=\nabla\times\vec{A}
\end{equation}
\begin{equation}
\label{energy}
\frac{\partial e}{\partial t}+\vec{\nabla}\cdot[(e+p)\vec{v}]=\rho\vec{g}\cdot\vec{v},
\end{equation}
where $t$ is time, $\rho$ is density, $\vec{v}$ velocity, $p$ thermal pressure, and $\vec{B}$ the magnetic field.
The total energy density $e$ is given by
\begin{equation}
\label{enercouple}
e=\frac{p}{\gamma-1}+\frac{1}{2}\rho\vec{v}^2+\frac{\vec{B}^2}{8\pi}
,\end{equation}
where $\gamma=5/3$ denotes the ratio of specific heats.
The expression for solar gravitational acceleration is
\begin{equation}
\label{solargravity}
\vec{g}=-\frac{G M_{\odot}}{r^2}\hat{r},
\end{equation}
where $G$ is the gravitational constant, $M_{\odot}$ denotes the mass of the Sun,
$r$ is the radial distance from the centre of the Sun,
and $\hat{r}$ is the corresponding unit vector.
In order to gain accuracy in the description of the thermal pressure,
we make use of the magnetic field splitting technique \citep{Powell1999},
as explained in detail in Sec.2.3 of \citet{Pagano2013a}.

The boundary conditions are treated with a system of ghost cells
and match those used in \citet{MackayVanBallegooijen2006A}.
Open boundary conditions are imposed at the outer boundary, reflective boundary conditions are set at the $\theta$ boundaries,
and the $\phi$ boundaries are periodic.
The $\theta$ boundary condition does not allow any plasma or magnetic flux through,
while the $\phi$ boundary condition allows the plasma and magnetic field to freely evolve across the boundaries.
In our simulations the expanding and propagating flux rope only interacts with the $\theta$ and $\phi$ boundaries
near the end of the simulation, thus they do not affect our main results
regarding the initiation and propagation of the CME.
At the lower boundary we impose a fixed boundary condition taken from the first four $\theta$-$\phi$
planes of cells derived from the GNLFFF model.
The computational domain is composed of $256\times128\times128$ cells, distributed on a uniform grid.
The simulation domain extends over $3$ $R_{\odot}$ in the radial direction starting from
$r=R_{\odot}$. The colatitude, $\theta$, spans from $\theta=30^{\circ}$ to
$\theta=100^{\circ}$ and the longitude, $\phi$, spans $90^{\circ}$.
The simulated domain extends to a larger radial distance than the domain used in \citet{MackayVanBallegooijen2006A}
from which we import the magnetic configuration.
To define the magnetic field for $r>2.5$ $R_{\odot}$, we assume it to be purely radial ($B_{\theta}=B_{\phi}=0$) where
the magnetic flux is conserved:
\begin{equation}
\label{brover25r}
B_r(r>2.5 R_{\odot},\theta,\phi)=B_r(2.5 R_{\odot},\theta,\phi)\frac{2.5^2}{r^2}.
\end{equation}
We set the maximum value of the magnetic field intensity to be $B_{max}=63$ $G$ in the present simulation.

\begin{figure}[!htcb]
\centering
\includegraphics[scale=0.32]{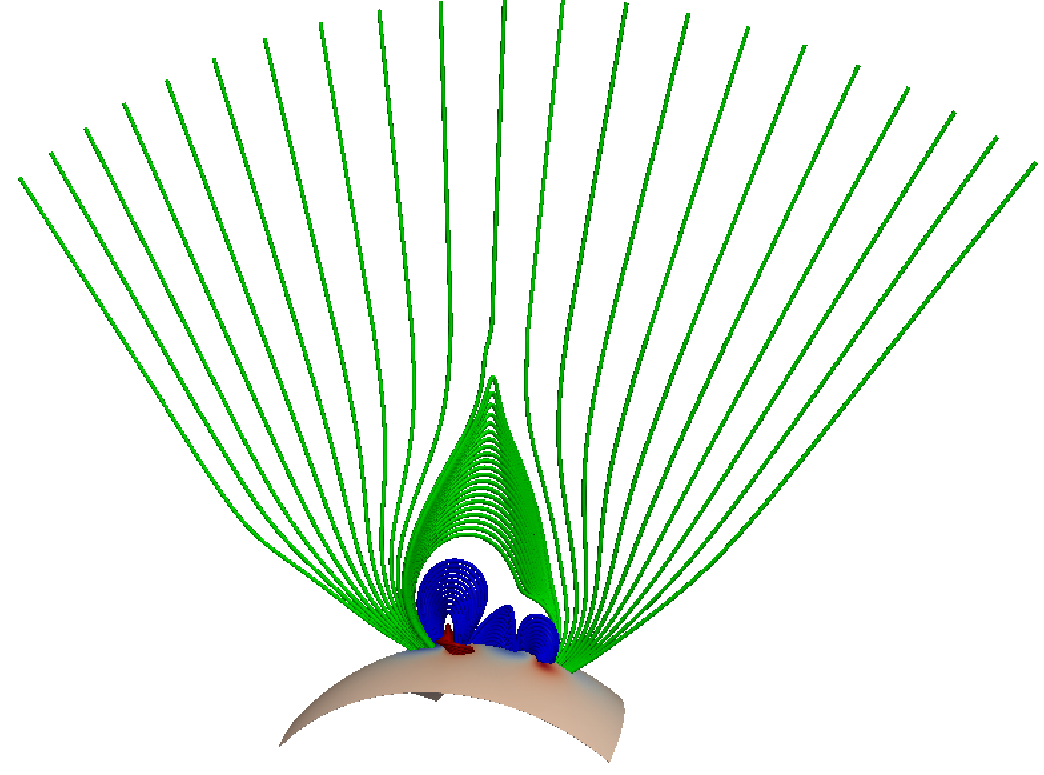} 
\caption{Magnetic field configuration used as the initial condition in all the MHD simulations.
Red lines represent the flux rope, blue lines the arcades, green lines the external magnetic field.
The lower boundary is coloured according to the polarity of the magnetic field
from blue (negative) to red (positive) in arbitrary units.}
\label{initialmagnetic}
\end{figure}

Figure \ref{initialmagnetic} shows a 3D plot of the initial magnetic configuration,
which is imported from Day 19 of the simulation by \citet{MackayVanBallegooijen2006A}.
The flux rope (red lines) lies in the $\theta$ direction.
The flux rope is close to the point where an eruption will occur,
as it can no longer be held down by the overlying arcades.
The arcades are shown by the blue lines, above which lies the
external magnetic field lines (green lines).
Some of the external magnetic field lines belong to the external arcade while others are open.
For the plasma properties we prescribe a non-isothermal solar corona,
with a simplified version of the treatment described in \citet{Pagano2014}.
To produce the temperature distribution we use the function $T(\vec{B})$,
\begin{equation}
\label{temperaturedistr}
T(\vec{B})=\left[\frac{6}{2+(B_{\theta}/|B|)}-2\right](T_{out}-T_{min})+T_{min}
,\end{equation}
where $T_{out}$ is the value of $T(\vec{B})$ when $B_{\theta}=0$. 
The parameter $T_{min}$ determines the minimum allowed value for $T(\vec{B})$ and occurs where $B_{\theta}=|B|$.
While the choice of this ad hoc analytic formula may seem strange,
it is justified by the fact that in our set up the flux rope lies in the $\theta$ direction
with a positive $B_{\theta}$.
It is the only structure with a strong magnetic shear in the initial condition of our simulation.
The form applied in Eq.\ref{temperaturedistr} allows us to produce
a cool dense region at the location of the flux rope (i.e. high $B_{\theta}$).
In principle it is possible to generalize this temperature distribution
by replacing the $\theta$ direction with
the direction of the flux rope axis.
However, for the present simulations this would have little effect.

The distribution of thermal pressure is independently specified using the solution for hydrostatic equilibrium
with a uniform temperature set equal to $T_{out}$,
\begin{equation}
\label{pressurestratification}
p=\frac{\rho_{LB}}{\mu m_p}k_B 2 T_{out} \exp\left({-\frac{M_{\odot}G \mu m_p}{2 T_{out} k_B R_{\odot}}}\right) \exp\left({\frac{M_{\odot}G \mu m_p}{2 T_{out} k_B r}}\right),
\end{equation}
where $\rho_{LB}$ is the density at $r=R_{\odot}$ when $|B|=0$,
$\mu=1.31$ is the average particle mass in the solar corona,
$m_p$ is the proton mass,
$k_B$ is Boltzmann constant.
Finally, the density is simply given by the equation of state applied to Eq.\ref{temperaturedistr} and Eq.\ref{pressurestratification}:
\begin{equation}
\label{eos}
\rho=\frac{p}{T(\vec{B})}\frac{\mu m_p}{k_B}
.\end{equation}

\begin{figure}[!htcb]
\centering
\includegraphics[scale=0.25]{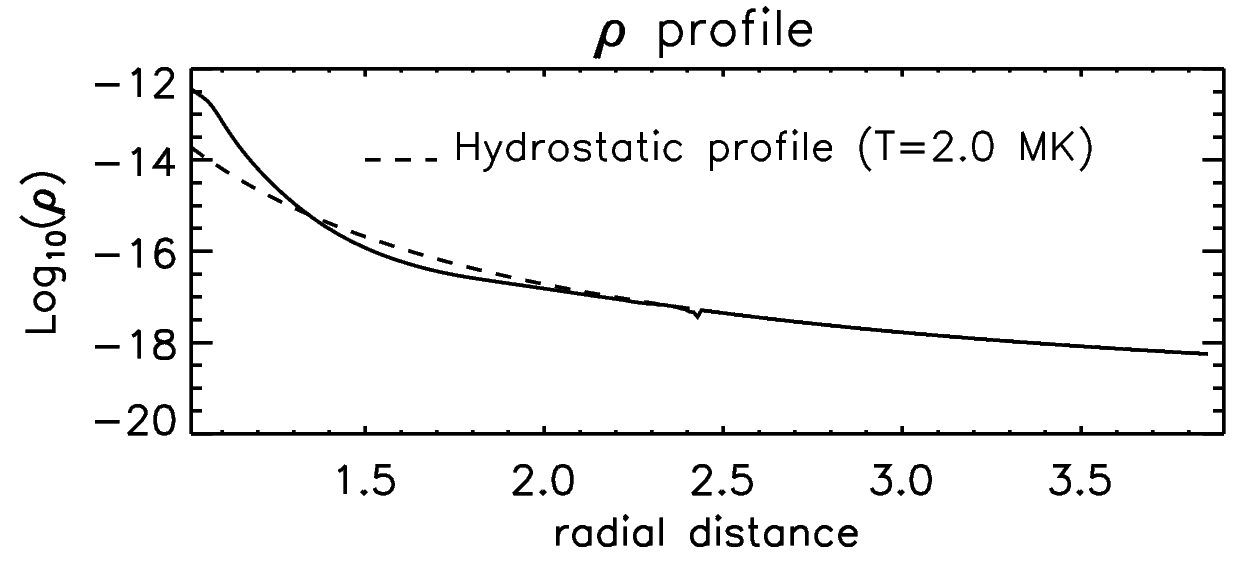} 
\caption{Profiles of $Log_{10}(\rho)$ along the radial direction above the centre of the LHS bipole
at $t=0$ $min$. The dashed lines show the equivalent profiles for 
a coronal atmosphere in hydrostatic equilibrium at $T=2$ $MK$.}
\label{ptrhoprofile}
\end{figure}

In our simulation we choose
$T_{out}=2$ $MK$, 
$T_{min}=0.1$ $MK$, and
$\rho_{LB}=3.5\times10^{-14}$ $g/cm^3$.
With these values, we obtain the atmospheric profile shown in Fig.\ref{ptrhoprofile}; this figure {} draws a radial cut of density (solid line) from
the lower boundary to the external boundary passing through the centre of the LHS bipole
(where the flux rope lies).
As the radial distance increases,
the coronal density profile decreases
as a result of  gravitational stratification.
Below $1.3$ $R_{\odot}$ there is an excess of density compared to the hydrostatic profile (dashed line).
This is due to the presence of the flux rope that is denser and cooler than the surroundings,
as prescribed by Eq.\ref{temperaturedistr}.
The jump at $2.5$ $R_{\odot}$ in the profile of density is a result of the transition
between the magnetic field configuration interpolated from the GNLFFF model ($r<2.5$ $R_{\odot}$)
and the perfectly radial magnetic field ($r>2.5$ $R_{\odot}$). 
Here a small $B_{\theta}$ component of the magnetic field leads to a visible effect in the density profile.
The dynamic effect of this jump is negligible,
as the coronal dynamics is dominated by the ejection of the flux rope.
In particular, the density at this location is negligible in comparison with the density of the ejected flux rope
\citep[See the Appendix in][]{Pagano2013b}.

\subsection{Evolution of the MHD simulation}
\label{mhdevolution}
\begin{figure}[!htcb]
\centering
\includegraphics[scale=0.22]{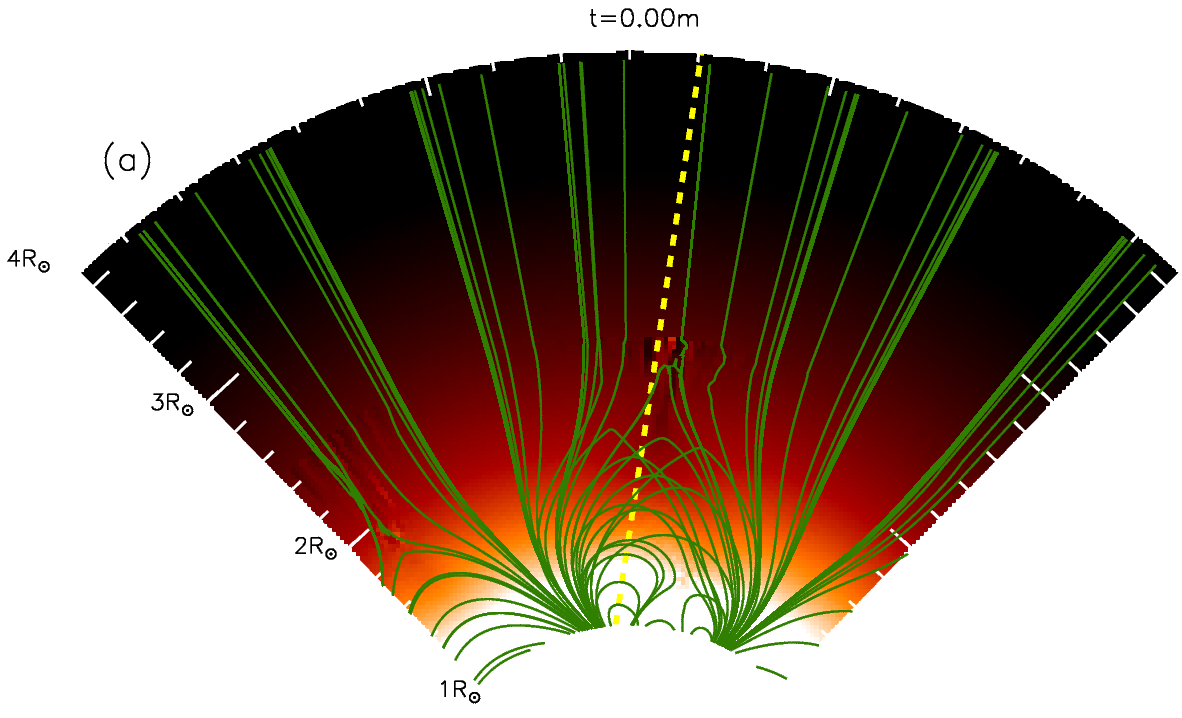} 
\includegraphics[scale=0.16]{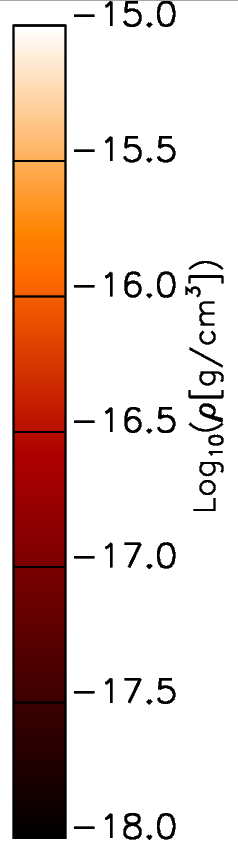} 

\includegraphics[scale=0.22]{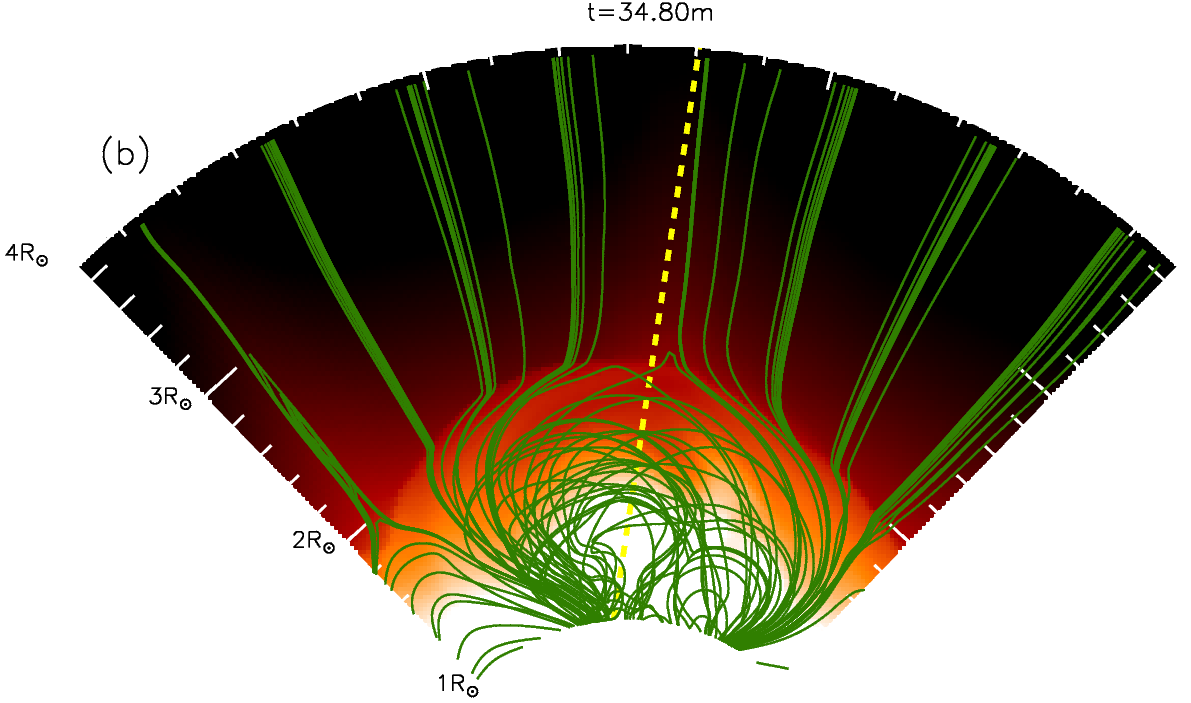} 
\includegraphics[scale=0.16]{fig03s.png} 

\includegraphics[scale=0.22]{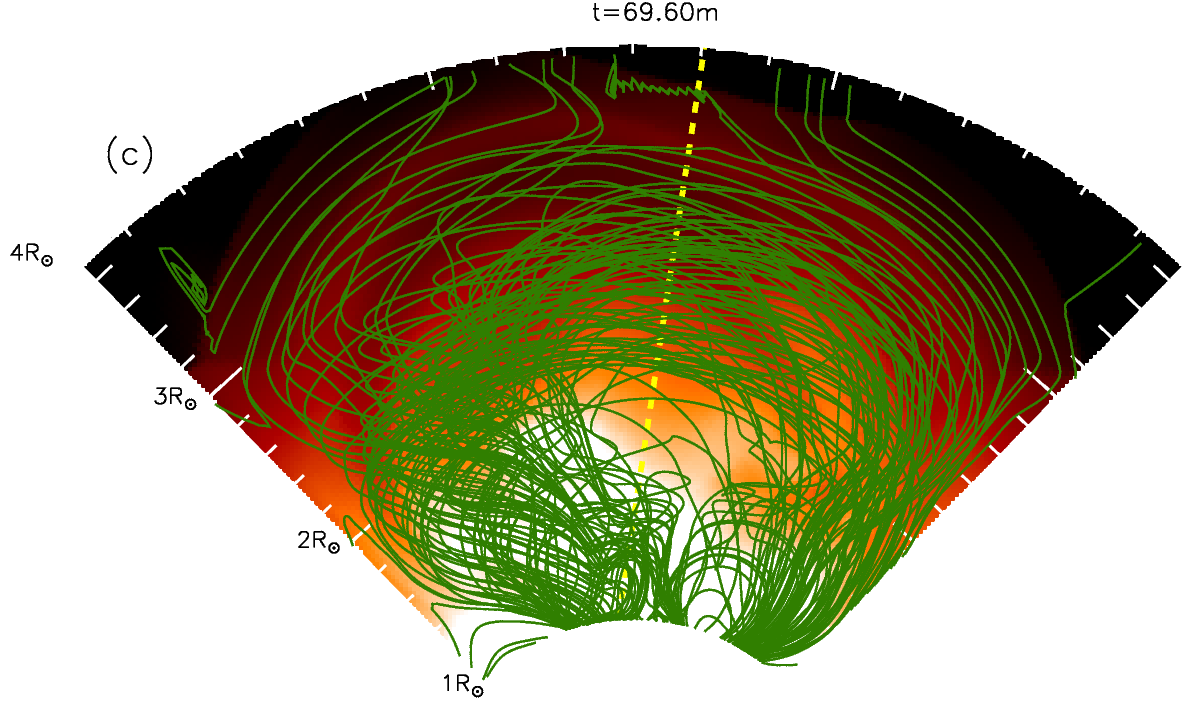} 
\includegraphics[scale=0.16]{fig03s.png} 
\caption{Maps of $Log_{10}(\rho)$ in the $(r-\phi)$ plane
passing through the centre of the bipoles at $t=0$, $t=34.80$, and $t=69.40min$.
Superimposed are magnetic field lines plotted from the same starting points (green lines).
The yellow dashed line shows the flux rope propagation line.}
\label{evolT2B500cdFR}
\end{figure}

The MHD simulation shows a very similar evolution to that found in our previous work.
The evolution is illustrated in Fig.\ref{evolT2B500cdFR} where maps of density
are shown in the ($r-\phi$) plane, passing through the centre of the bipoles.
In each image  we also superimpose magnetic field lines.
Initially the flux rope lies near the lower boundary
and as soon as the system is allowed to evolve it is ejected upwards.
The cause of the ejection is explained in \citet{Pagano2013a}
where we describe how the magnetic stress built up
during the flux rope formation is released.
Once the ejection occurs, a high density structure rises and expands.
Initially, the high density region maintains the shape of the ejected flux rope (Fig.\ref{evolT2B500cdFR}b);
however, near the end of the simulation it is less identifiable with the flux rope (Fig.\ref{evolT2B500cdFR}c).
After approximately 100 minutes, the high density region reaches the outer boundary at $4$ $R_{\odot}$.
It should be noted that the flux rope propagates non-radially
(yellow dashed line in Figs. \ref{evolT2B500cdFR}a-c)
in the direction of the null point, which lies above the arcade system.
It follows the non-radial path as the confining Lorentz force exerted by the magnetic arcades is weakest there 
and so this becomes a favourable escape direction.
This leads to a deflection of the flux rope by a few degrees from a perfectly radial propagation.
The magnetic field configuration undergoes a major evolution and reconfiguration as a result of the ejection.
Magnetic flux is expelled outwards and at $t=69.60$ $min$ (Fig.\ref{evolT2B500cdFR}c)
a region of compressed magnetic field is visible ahead of the ejection,
where the front of the ejection compresses both plasma and magnetic flux.

\begin{figure}[!htcb]
\centering
\includegraphics[scale=0.25]{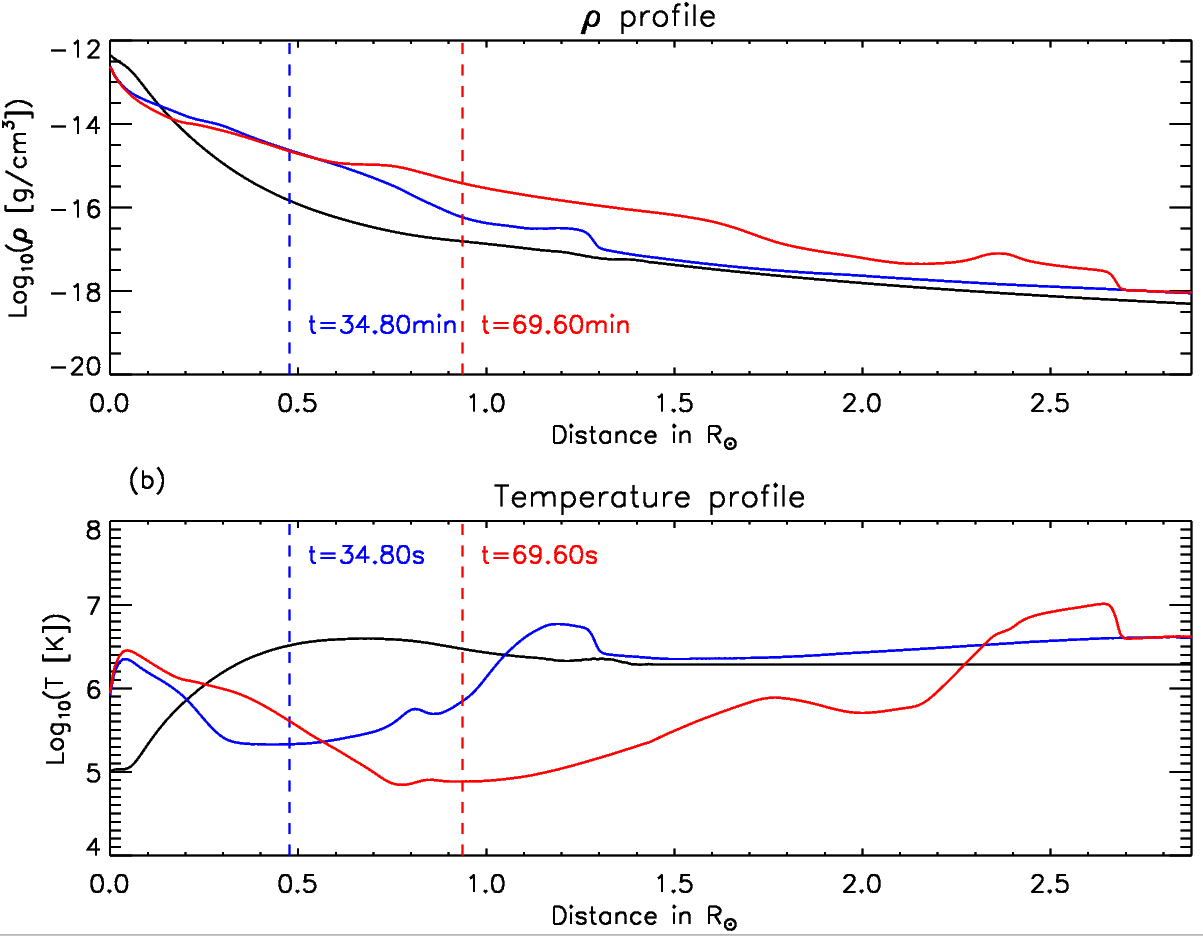} 
\caption{Profiles of $Log_{10}(\rho)$ along the yellow dashed line in Fig.\ref{evolT2B500cdFR},
which represents the direction of the flux rope ejection
at $t=0$ $min$ (black line), $t=34.80.60$ $min$ (blue line), and $t=69.60$ $min$ (red line).
The dashed lines mark the position of the centre of the flux rope at each time.}
\label{rhocuts}
\end{figure}

\begin{figure}[!htcb]
\centering
\includegraphics[scale=0.25]{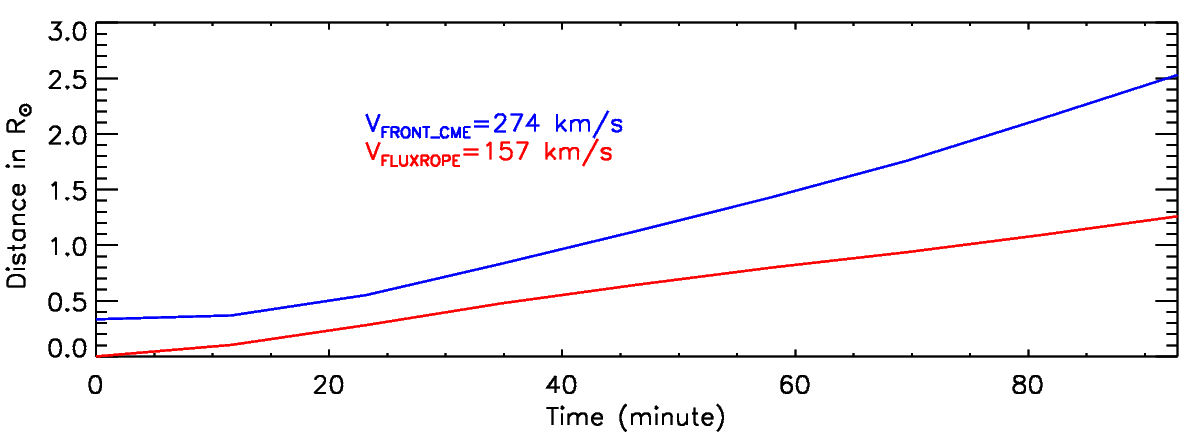} 
\caption{Position of the radial distance of the centre and top of the flux rope as a function of time.}
\label{centerfrposition}
\end{figure}

The flux rope ejection leads to an increase in density in the outer corona
that propagates outwards and can be labelled as a CME.
It also exhibits the three-part density structure
 associated with CMEs.
Figure \ref{rhocuts} shows the density profile along the
line of propagation of the flux rope at each of the times shown in Fig.\ref{evolT2B500cdFR}.
The dashed vertical lines show the position of 
the centre of the flux rope at each time.
In particular at $t=69.60$ $min$ we have a CME front where 
the density has increased by approximately two orders of magnitude
compared to the value of density at the same location at $t=0$.
Behind the front there is a void, where the density is about
half that found in the front.
Further behind the flux rope constitutes the core of the CME,
where the density is approximately one to two orders of magnitude
higher than that of the void.
Finally, Fig.\ref{centerfrposition} shows the position of the centre and top of the flux rope as a function of time
along the line of propagation.
The centre of the flux rope is given by the maximum of the quantity $B_{\theta}/|B|$ along the propagation line
and the top is located where the quantity $B_{\theta}/|B|$ changes sign above the centre.
The parameters we have chosen for the simulation give conditions in which 
the flux rope is expelled from the solar corona at a typical velocity.
The flux rope undergoes an acceleration phase for the first 30 minutes
and then it travels outwards at a constant speed.
The overall average speed of the flux rope is about $\sim157$ $km/s$,
whereas its front travels slightly faster at $\sim274$ $km/s$,
as it expands whilst propagating outwards.

\section{Synthesis of METIS observations in white light}
\label{metisobservation}
The synthesis of total white-light (tB) and polarized (pB) brightness images is carried out in 3 steps.
In step one we interpolate the outputs of the spherical 3D MHD simulation onto a 3D cartesian grid.
For the second step, we compute the total and polarized brightness of each plasma element
in the observer's direction. Finally, for the third step we integrate along the line of sight (LOS).

The first step is carried out through a 3D interpolation
that has already been successfully applied in \citet{Pagano2014}.
This interpolation is carried out to allow the integration along the LOS of the originally spherical domain, thus transforming the domain of the spherical MHD simulation into a cartesian grid.
In this grid the flux rope lies on the solar disk at about $24^\circ$
from the plane of sky (POS) in the direction of the observer.
This can be done as the longitudinal coordinate is arbitrary.
The choice of $24^\circ$ is made to study a general case where the CME is not ejected too close to nor too far away from the POS.
The resulting cartesian grid has an origin at the centre of the Sun,
$z$ is the direction along the LOS,
$y$ is the direction parallel to the NS axis of the Sun,
and the $x$ direction is perpendicular to both $z$ and $y$.
Owing to geometrical effects,
the spherical simulation cannot fill every point of the cartesian box with values.
Therefore, to complete the cartesian box
the initial density and temperature distribution
are mirrored about the planes:
$\phi=0^{\circ}$, $\phi=90^{\circ}$, 
$\theta=30^{\circ}$, and $\theta=100^{\circ}$.
Multiple mirrored copies of the plasma distribution for the MHD domain at $t=0$ are used
to populate part of the cartesian box that cannot be sampled from the spherical simulation. 
In such a way
i) plasma is present in the cartesian box everywhere  ($r\leq4$ $R_{\odot}$), and
ii) outside the domain specified by the spherical simulation,
the corona remains unperturbed as given by the initial condition.

\begin{figure}[!htcb]
\centering
\includegraphics[scale=0.5]{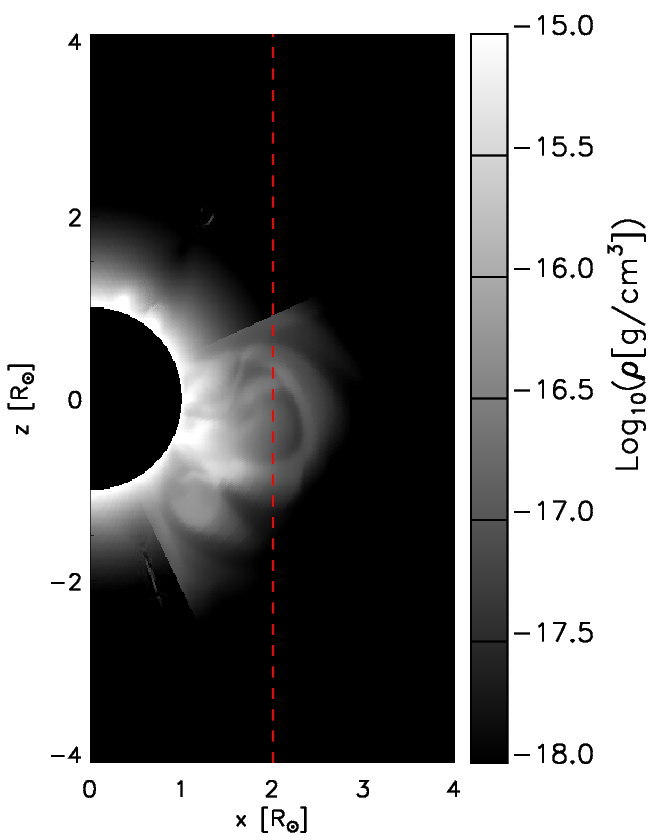} 
\caption{Map of $Log_{10}(\rho)$ in the cartesian box at $y=0.3$ $R_{\odot}$
with a LOS at $x=2$ $R_{\odot}$
overplotted with red dashed line.}
\label{losxzy}
\end{figure}

Figure \ref{losxzy} shows a generic LOS (red dashed line) drawn on
the corresponding density map on the $x-z$ plane at $y=0.3$ $R_{\odot}$ and $t=69.60$ $min$.
The flux rope ejection produces many distinct density structures travelling 
outwards and the LOS crosses some of them.
In contrast, the corona outside of the spherical MHD simulation domain
is unperturbed and remains in its initial state.

\begin{figure}[!htcb]
\centering
\includegraphics[scale=0.25]{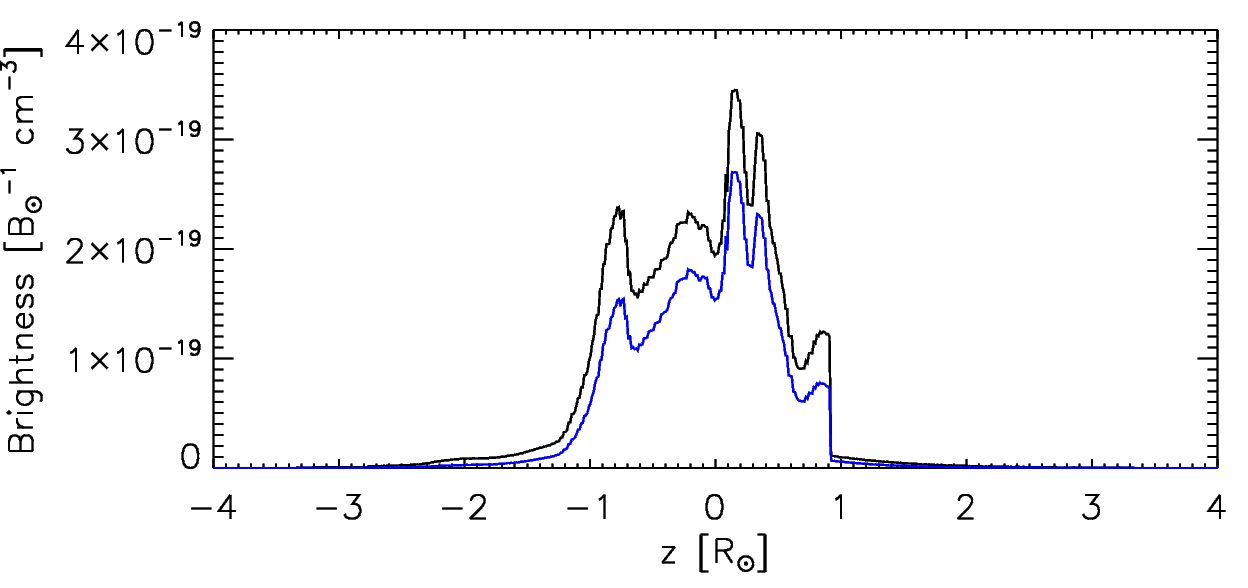} 
\caption{Emission in white light tB (black line) and pB (blue line) along the LOS shown in Fig.\ref{losxzy}.}
\label{losz}
\end{figure}

The second step is to compute the tB and pB emissions from each
plasma element of the cartesian box.
This is carried out by computing the Thomson scattered light from a single electron
in each cell of the cartesian box, depending on 
the impact distance of the LOS and
the angle from the plane of sky \citep{Minnaert1930}.
The contribution is then multiplied by the electron number density of each cell.
We point out here that the white-light emission of a plasma element
only depends on the plasma density, its position with respect to the solar disk, and the specific scattering angle.
Figure \ref{losz} shows the computed tB and pB emissions
along the LOS in Fig.\ref{losxzy}.
The vast majority of the emission is from the region within $1$ $R_{\odot}$
from the plane of sky ($z=0$).
Therefore, the assumptions made to populate
the peripheral regions of the cartesian box have no significant impact on our study.

The final step consists of integrating the contribution from each plasma element along a given LOS.
In the present work we use $320\times400$ lines of sight and 
we choose to show only the lines of sight crossing
the METIS field of view when Solar Orbiter reaches  perihelion at $0.29$ $AU$ from the Sun
($1.6 R_{\odot}<\sqrt{x^2+y^2}< 3.1 R_{\odot}$).

\begin{figure}[!htcb]
\centering
\includegraphics[scale=0.18]{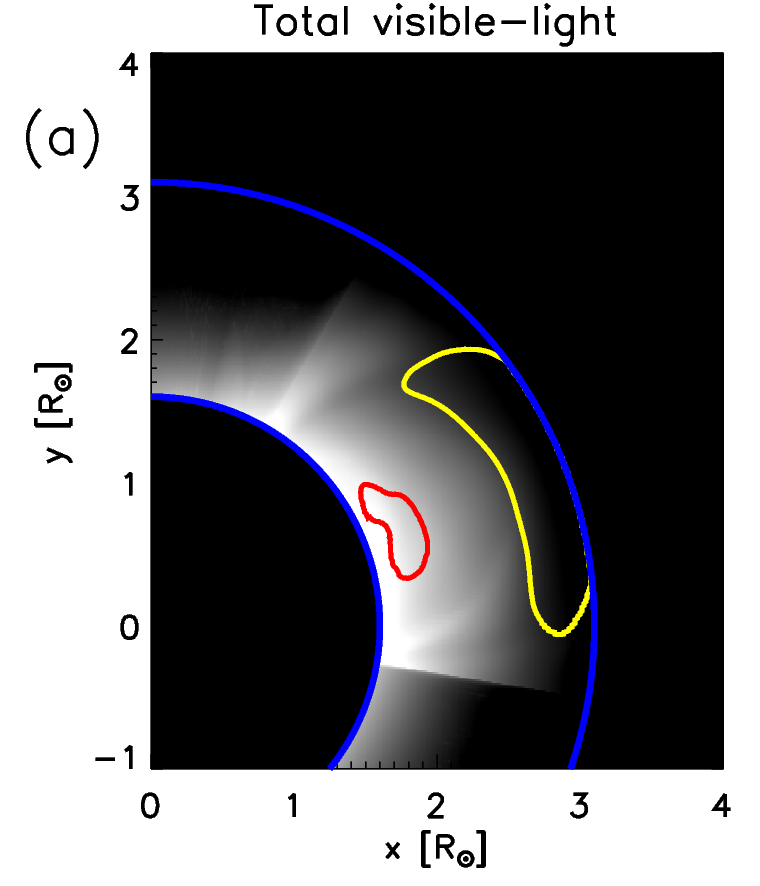} 
\includegraphics[scale=0.18]{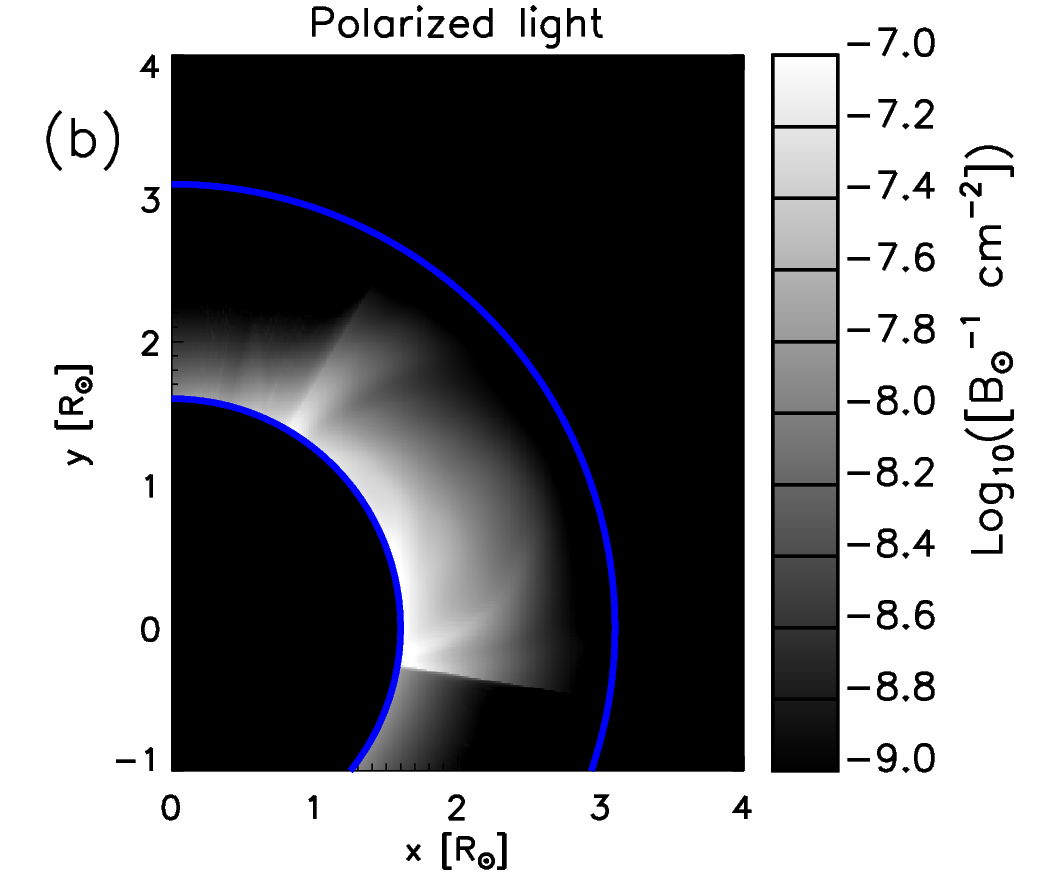} 
\caption{(a) Synthetic image of the total white-light brightness ($tB$)
integrated along the line of sight in the METIS FOV.
The red contour roughly identifies the flux rope position where the ratio
between the $y$ component of the magnetic field
integrated along the line of sight divided by the magnetic field intensity
is above the arbitrary positive threshold of 0.4,
and the yellow contour the position of the ejection front
where this quantity is lower than the arbitrary value of -0.43.
(b) Synthetic image of the polarized white-light brightness ($pB$)
integrated along the line of sight in the METIS FOV.
Both images are synthesized from the MHD simulation at $t=69.60$ $min$
and using a METIS field of view at  perihelion at $0.29$ $AU$ from the Sun and extending from
$1.6$ $R_{\odot}$ to $3.1$ $R_{\odot}$ (blue lines).}
\label{pol0060}
\end{figure}

Figure \ref{pol0060}a shows the synthesized observation of METIS in white-light from our MHD simulation at $t=69.60$ $min$,
where the red contour highlights the approximate location of the centre of the flux rope axis, and 
the yellow contour is the CME front.
The two blue lines mark the METIS field of view.
At this radial distance from the Sun, the flux rope is still well defined
and visible in the observations as the brightest region of the ejection.
METIS will observe close enough to the solar surface to catch 
coherent structures before they merge into the solar wind stream of the outer corona.
In contrast, the front of the CME seems quite faint in our case study.
Figure \ref{pol0060}b shows the polarized component of the white-light and
this mostly reproduces the same structures seen in Fig.\ref{pol0060}a,
except with lower brightness.
Both panels in Fig.\ref{pol0060} show a sharp discontinuity in emission at the $\theta$-borders
of the simulation domain.
This is due to the limited $\theta$ and $\phi$ extension of our spherical domain,
as in these regions we have assumed the corona to be maintained unperturbed.

\section{Results}
\label{result}

From the synthesis of  METIS observations, we obtain tB and pB images that are 
equivalent to actual white-light observations of the
solar corona with the METIS coronagraph.
In this section, we present the results obtained from
the analysis of synthesized images with polarization ratio and excess brightness techniques.
To derive the CME plasma 3D distribution and electron density,
we compare the results with the actual density distribution from the MHD simulation.
Finally, we interpret the comparison in terms of the analysis carried out in \citet{BemporadPagano2015}.

\subsection{Reconstruction of the 3D CME structure and propagation direction}
\label{foldedcentre}
We apply the polarization ratio technique to synthesized METIS difference images,
i.e. images from which the pre-CME quiet corona emission has been subtracted.
In observations, this subtraction is important in order to isolate the so-called excess brightness
(i.e. the total brightness in white light due to the CME plasma alone tB$_{CME}$).
It removes the spurious emission due to background coronal plasma located
along the same LOS but external to the CME volume.
In this study, the white-light image corresponding to the MHD simulation at $t=0$ is used as the quiet corona,
thus our assumption of the density distribution
outside of the spherical simulation becomes irrelevant for the results of the analysis.

\begin{figure}[!htcb]
\centering
\includegraphics[scale=0.20]{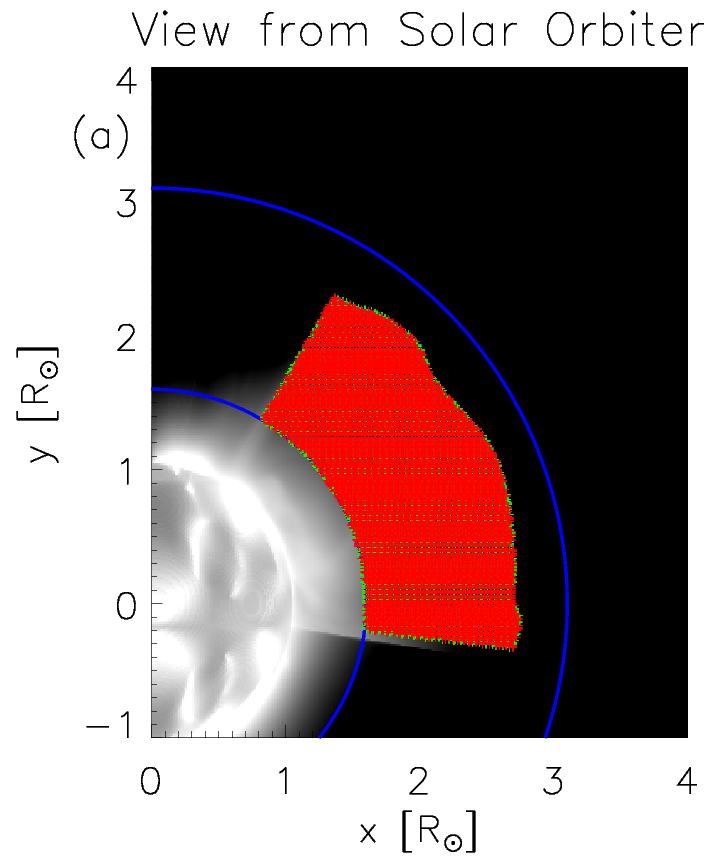} 
\includegraphics[scale=0.20]{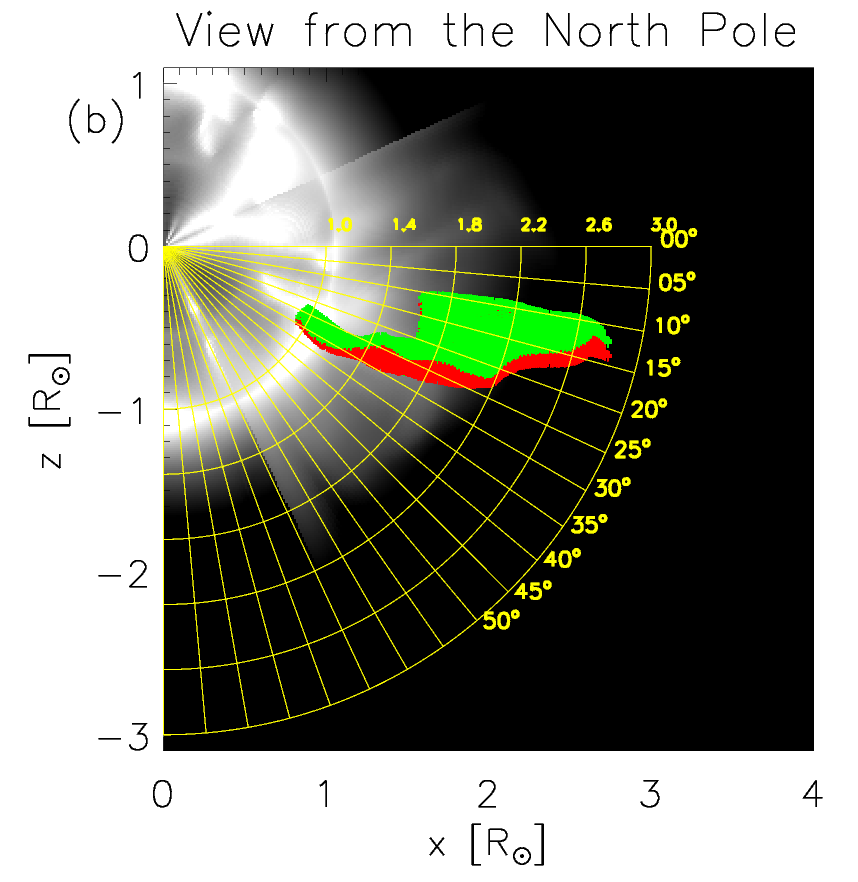} 
\includegraphics[scale=0.20]{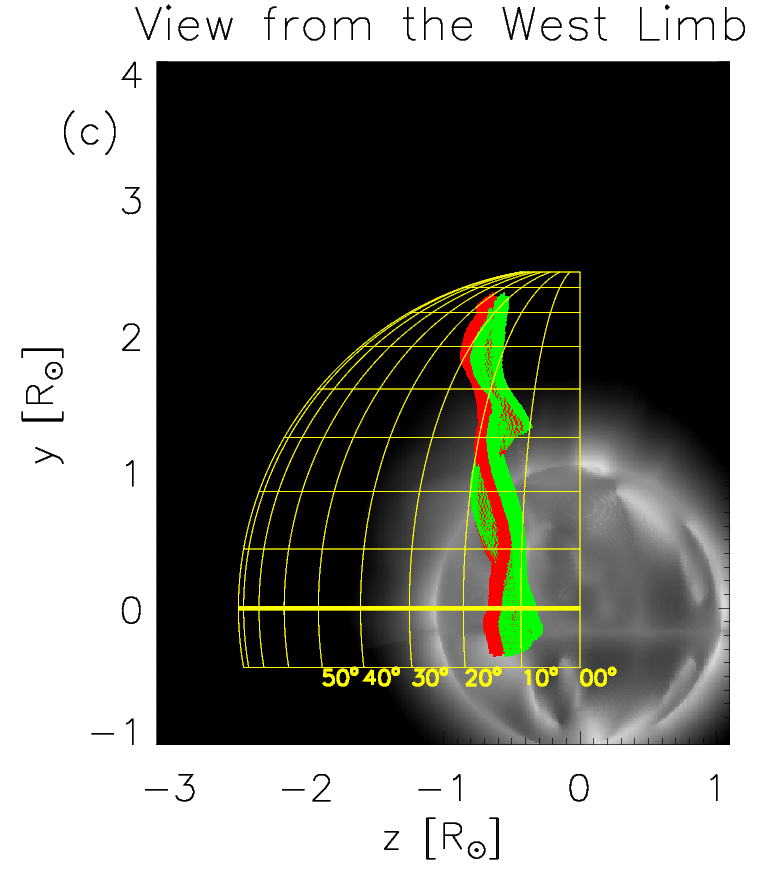} 
\includegraphics[scale=0.20]{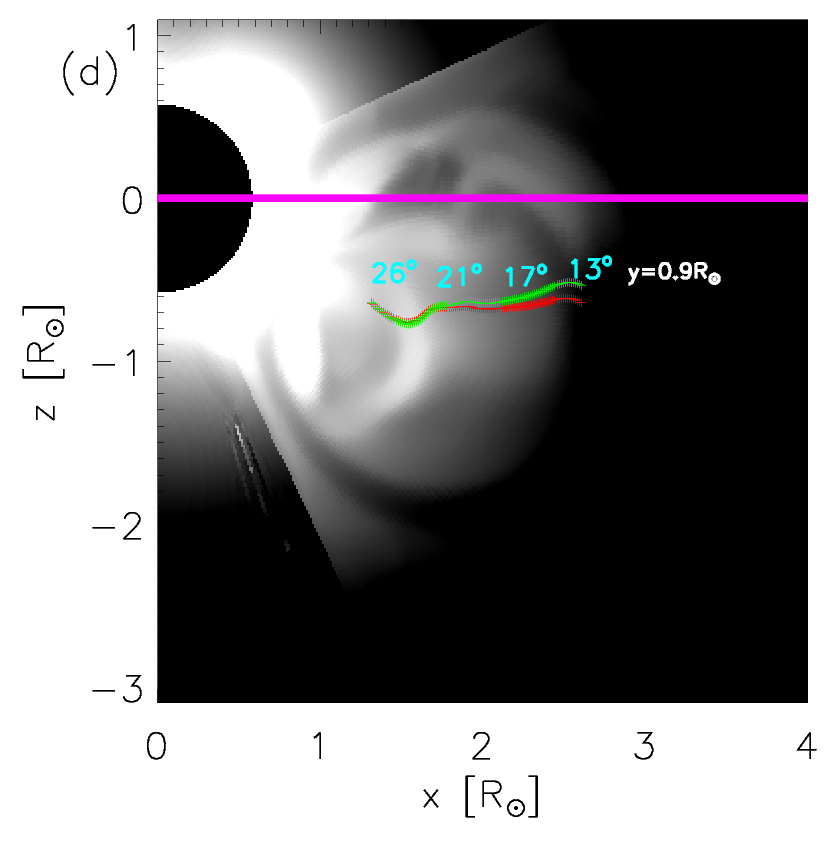} 
\caption{Maps of $Log_{10}$ of column density
(range between $10^{-15}$ $g/cm^2$ and $10^{-12}$ $g/cm^2$) seen from three different points of view; 
 superimposed is the cloud of points inferred from the polarization ratio technique (red crosses)
and the cloud of point computed from Eq.\ref{foldedcom} (green crosses) projected onto the plane of sky:
(a) point of view parallel to direction $z$ and $(x,y)$ coordinates of clouds of points;
(b) point of view parallel to direction $y$ and $(x,z)$ coordinates of clouds of points;
(c) point of view parallel to direction $x$ and $(z,y)$ coordinates of clouds of points.
(d) Maps of white-light emission seen from a point of view
parallel to the $y$ direction;  superimposed is
a subset of points from the two clouds at $y=0.9$ $R_{\odot}$.
In light blue we give the angle with the plane of sky for some of the points.
The thick magenta line marks the plane of sky.}
\label{cloudszyx}
\end{figure}

We identify a sample of lines of sight to apply the polarization ratio technique.
Figure \ref{cloudszyx}a shows the identified lines of sight,
all intercepting our MHD simulation box and
showing a white-light brightness above a given threshold.
These lines of sight identify a distinct set of $(x,y)$ points.
For each (x,y) point the polarization ratio technique gives
a corresponding position $z$ of the emitting plasma along the LOS,
thus a cloud of points (x,y,z).
This cloud of points is compared with the cloud of points 
corresponding to the centre of mass of the density distribution folded about the plane of sky
(called hereafter folded centre of mass) \citep{BemporadPagano2015}:

\tiny 
\begin{equation}
\begin{split}
CM_{folded}(x,y)= \\
\frac{\int_{\infty}^{0} z[(\rho(z,x,y,t)-\rho(z,x,y,0))+(\rho(-z,x,y,t)-\rho(-z,x,y,0))] \, dz}
{\int_{\infty}^{0} (\rho(z,x,y,t)-\rho(z,x,y,0))+(\rho(-z,x,y,t)-\rho(-z,x,y,0)) \, dz}
\label{foldedcom}
\end{split}
.\end{equation}
\normalsize 
Thus we have a cloud of points $(x,y,z)$ resulting from the application
of the polarization ratio technique (red crosses hereafter)
and a second cloud of points $(x,y,z)$ (green crosses hereafter)
computed from the positions of the folded centre of mass along each LOS in the cartesian simulation box.
The two clouds are constructed in a way to have the same $(x,y)$ values, but can differ in $z$ values.
Figure \ref{cloudszyx}a shows the position of these points
as seen by the observer.
In Fig.\ref{cloudszyx}b we show the two clouds of points projected
onto the $x-z$ plane (corresponding to a view from the north pole),
and in Fig.\ref{cloudszyx}c we show the two clouds of points projected
onto the $z-y$ plane (corresponding to a view perpendicular to the equatorial axis and the observers point of view).
In all plots, red crosses represent the cloud of points obtained from the polarization ratio
and the green crosses represent the cloud of points obtained from Eq.\ref{foldedcom}.
As explained in \citet{BemporadPagano2015}, the position of the folded centre of mass
is the best approximation of the position output obtained from the polarization technique.
This is mainly due to the ambiguity in the polarization ratio expression
where it is not possible to distinguish whether the emitting plasma is located in front or behind the plane of the sky (POS).
In the following discussion we estimate the error associated with this approximation.

The polarization ratio technique catches
the 3D cloud of points of the folded centre of mass
with sufficient accuracy to describe the macroscopic structure, as the two clouds of points mostly overlap.
Also, the shapes are nearly identical
from both points of view perpendicular to the LOS (Fig.\ref{cloudszyx}b, Fig.\ref{cloudszyx}c).
Moreover, from Fig.\ref{cloudszyx}b we see that
both clouds of points are aligned along a direction of $25^{\circ}$ from the plane of sky
below $r=1.6$ $R_{\odot}$ and the clouds become more dispersed around a direction of $15^{\circ}$ from the plane of sky.
As can be seen in Fig.\ref{cloudszyx}b and Fig.\ref{cloudszyx}c, the two clouds are nearly identical in shape and position.
However, there is an offset between the two clouds, as the 
cloud of points obtained from the polarization ratio (red crosses) is about $2^{\circ}$ farther from the POS
than the cloud obtained from the folded centre of mass.
Figure \ref{cloudszyx}c shows that the two clouds follow this pattern independently of latitude.

A more detailed comparison can be carried out to measure the errors.
We consider a smaller sample of lines of sight,
all lying on the same $y$ coordinate.
Figure \ref{cloudszyx}d shows a set of points taken 
at $y=0.9 R_{\odot}$ (crossing the centre of the flux rope)
and in the image we indicate  the respective angle
with the plane of sky for a sample of representative points.
Green and red crosses never depart from one another by more than $0.1$ $R_{\odot}$
where the distance is the greatest at the outer boundary of the METIS field of view.
At the outer boundary the angle with the plane of sky is 
lower than $20^{\circ}$, as explained in \citet{BemporadPagano2015}.
This leads to an overestimation of the distance from the plane of sky
of the folded centre of mass position
(red crosses are farther from the plane of sky than green crosses).
Instead, at the location of the flux rope (brightest features)
the agreement between the points obtained by the polarization technique
and the folded centre of mass is much better.
In this region the structures lie at an angle of $\sim20^{\circ}$ or more with the plane of the sky.
As described in \citet{BemporadPagano2015}
the polarization ratio technique slightly underestimates the distance from the plane of sky
in such circumstances (green crosses are farther from the plane of sky than red crosses).
However, this effect is not as evident as in \citet{BemporadPagano2015}
owing to the presence of multiple structures along a line of sight,
which marks a significant difference from the idealized situation of a single blob.

\begin{figure}[!htcb]
\centering
\includegraphics[scale=0.33]{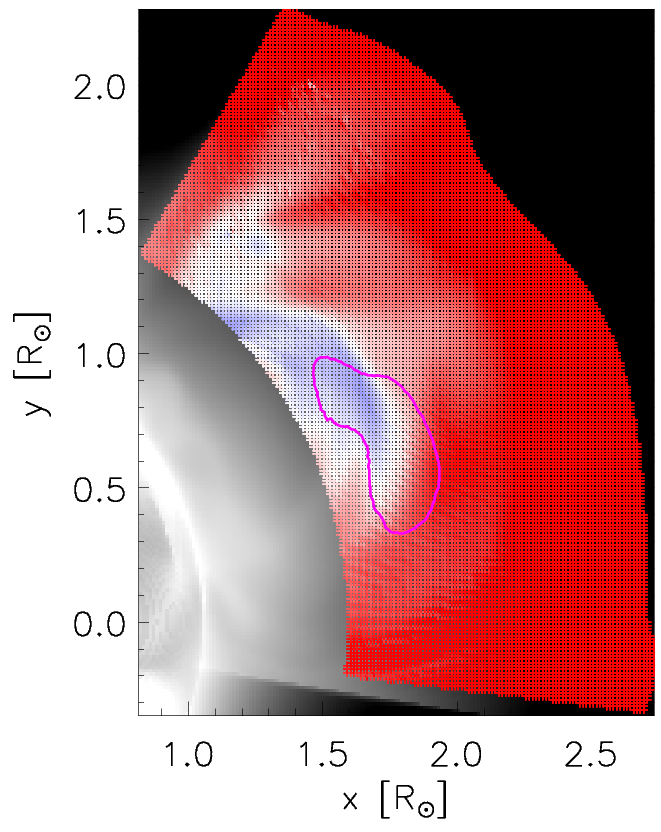} 
\includegraphics[scale=0.35]{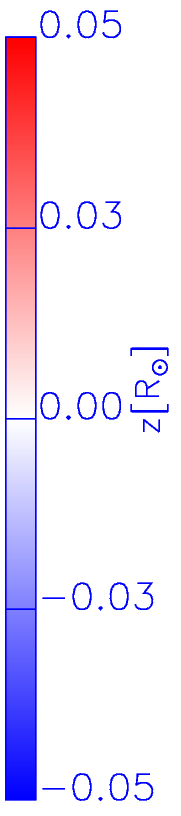} 
\caption{Map of differences of $z$ coordinates in $R_{\odot}$ for the two clouds of points,
overplotted on a map of $Log_{10}$ of column density as in Fig.\ref{cloudszyx}a.
Red regions correspond to where
the distance from the plane of sky computed from the polarization ratio technique is overestimated 
and blue regions correspond to where it is underestimated.
The magenta contour roughly identifies the flux rope position where the ratio
between the $y$ component of the magnetic field
integrated along the line of sight divided by the magnetic field intensity
is above the arbitrary positive threshold of 0.4.
}
\label{clouddiff}
\end{figure}
Figure \ref{clouddiff} shows the differences in the $z$ coordinate between the two clouds of points.
The differences in the position can be up to $\sim0.1$ $R_{\odot}$; however, at the location of the flux rope
we find that the differences are much smaller ( $<0.03$ $R_{\odot}$).
This is due to the increased accuracy of the polarization ratio technique
when the source of the white-light emission is highly localized.
Over the whole field of view,
the polarization ratio technique tends to overestimate the calculated distance
from the plane of sky compared to the position of the folded centre of mass.
In contrast it should be noted that the distance from the plane of sky is underestimated
in the region where the flux rope is present.
This occurs because in this region the folded centre of mass lies at more than $20^{\circ}$ from the POS
(see Fig.\ref{cloudszyx}b and Fig.\ref{cloudszyx}d)
and the measurement from the polarization ratio leads to an underestimation of the distance from the POS.

\subsection{Comparison with the centre of mass}
\label{truecentre}

In Sect.\ref{foldedcentre} we  illustrate that the polarization ratio technique
reconstructs the position of the folded centre of mass,
but introduces a measurement 
error of the order of $\sim0.03$ $R_{\odot}$.
However, an outstanding question is how these measurements compare with the
true position of the CME, i.e. the centre of mass along the LOS.

\begin{figure}[!htcb]
\centering
\includegraphics[scale=0.20]{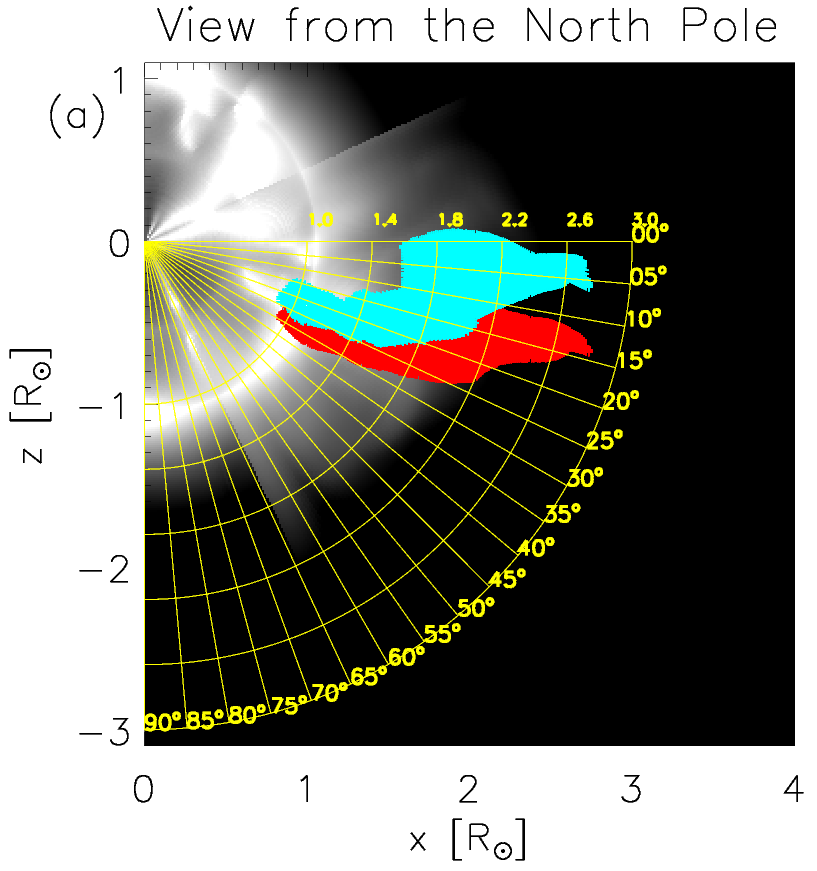} 
\includegraphics[scale=0.20]{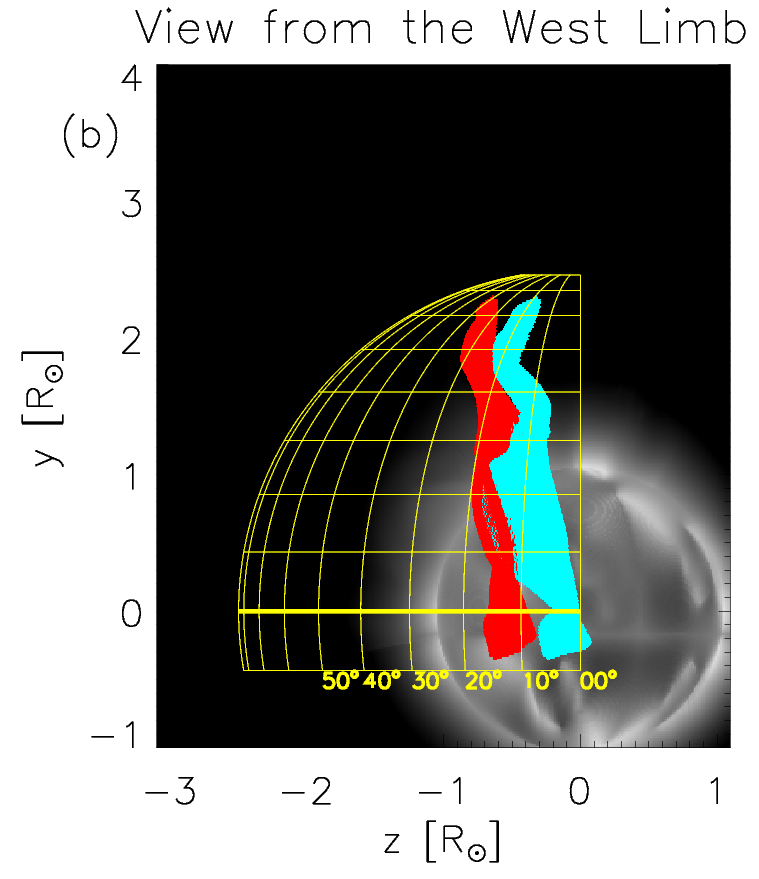} 
\caption{Maps of $Log_{10}$ of column density
(range between $10^{-15}$ $g/cm^2$ and $10^{-12}$ $g/cm^2$)
seen from two different points of view;  superimposed are the cloud of points inferred from the polarization ratio (red crosses)
and the cloud of points of the centre of mass along the LOS (light blue crosses) projected on the plane of sky:
(a) point of view parallel to the $y$ direction and $(x,z)$ coordinates of the clouds of points;
(b) point of view parallel to the $x$ direction and $(z,y)$ coordinates of the clouds of points.}
\label{cloudszyxnofolded}
\end{figure}

In Fig.\ref{cloudszyxnofolded} we show the images equivalent to
Fig.\ref{cloudszyx}b-c where we  compare the cloud of points
obtained by the polarization ratio technique (red crosses)
with the cloud of points determined from the centre of mass (light blue points).
The two clouds of points in Fig.\ref{cloudszyxnofolded}
show a very similar shape and extension;
however, they have a systematic offset,
significantly larger than the one highlighted in Fig.\ref{cloudszyx}.
In particular, the polarization ratio technique
systematically overestimates the distance from the plane of sky of the centre of mass
and the overestimation seems greater at larger radial distance 
from the Sun (Fig.\ref{cloudszyxnofolded}a).
For instance, this is evident for the cloud points below $1.5$ $R_{\odot}$
that correspond to the location of the flux rope.
The light blue points (centre of mass)
lie between $20^{\circ}$ and $25^{\circ}$,
which is where we have positioned the flux rope in our cartesian box.
In contrast, the red points (output of the polarization technique) lie between $25^{\circ}$ and $30^{\circ}$,
hence the distance from the POS is overestimated.

No clear dependence on latitude is seen (Fig.\ref{cloudszyxnofolded}b).
In this study the two clouds depart by about $10^{\circ}$,
but this value strongly depends on the amount of CME plasma
that is present beyond the POS.
This is a direct consequence
of the initial ejection position of the flux rope as well as its expansion.
In general the matching between the two clouds is 
significantly worse than in Fig.\ref{cloudszyx}
and the offset between the clouds in Fig.\ref{cloudszyxnofolded}
is $\sim0.2$ $R_{\odot}$.

\subsection{Electron density estimate and comparison with MHD simulation}
\label{densityestimate}
Another application of white-light images is to measure the density of the plasma of the CME.
The electron density has been computed (as  is usually done for real observations) from the excess brightness tB$_{CME}$ observed inside the simulated CME. This quantity is given by an integral of the white-light brightness due to all the electrons aligned along the considered LOS; the electron column density $N_e$ (cm$^{-2}$), corresponding to the LOS integral of the electron density distribution $n_e$ (cm$^{-3}$), is thus provided by dividing the tB$_{CME}$ by the brightness of a single electron tB$_e$.
However, owing to the geometry of Thomson scattering, the latter quantity depends on the location of the considered electron along the LOS;
in particular, tB$_e$ maximizes at the POS and then progressively decreases
when moving away from the POS along the LOS.
The usual assumption in the data analysis is that the bulk of the emission comes from the plasma located around the POS
(as also shown in Fig.\ref{losz}).
Therefore a lower limit estimate of the column density $N_e$ is obtained
by dividing the excess brightness by the tB$_e$ exactly on the POS ($z=0$).
We now explore a possible refinement of this technique:
we estimate the column density $N_e$ by dividing the excess brightness tB$_{CME}$
by the single electron brightness tB$_e$ assuming
that it is located on the position along the LOS determined from the polarization ratio technique.
Hence, we have two slightly different measurements of the column density.
First, by assuming that the white-light emission originates from a single bulk of plasma
placed either in the POS or
second, at the position along the LOS computed by the polarization ratio technique.
Such assumptions are reasonable in order to achieve an estimation of the plasma involved in a CME.
However, we need to verify that they can reproduce an actual map of the column density in a given field of view.

\begin{figure}[!htcb]
\centering
\includegraphics[scale=0.20]{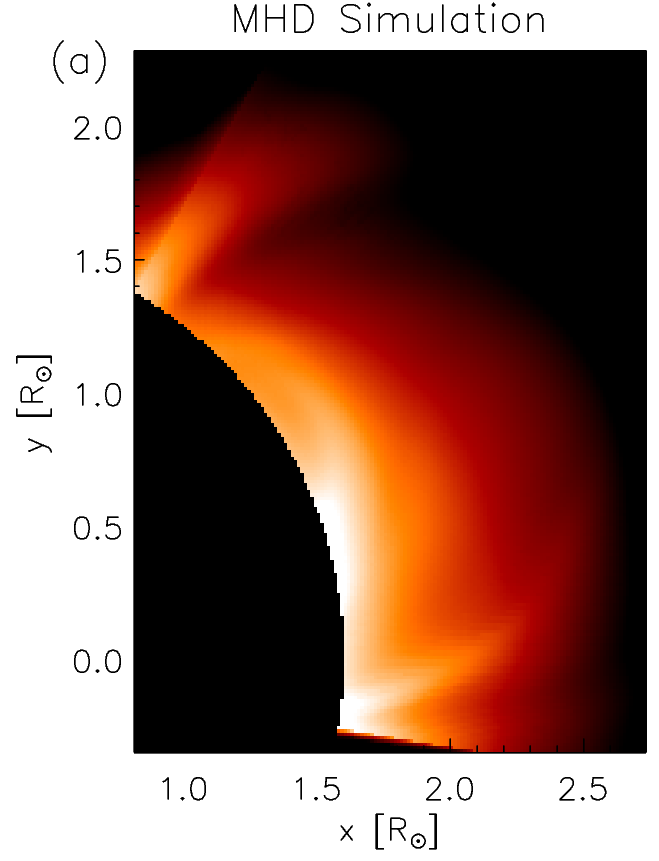} 
\includegraphics[scale=0.20]{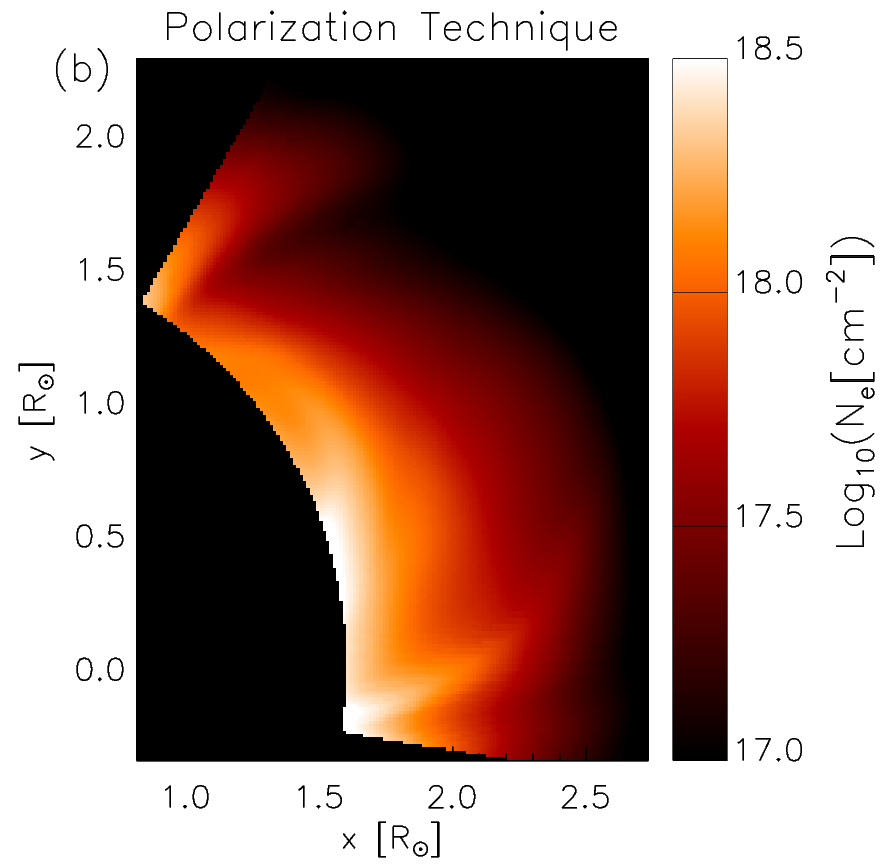} 
\caption{(a) Maps of $Log_{10}$ of column density number ($N_e$) computed from the MHD simulation.
(b) Maps of $Log_{10}$ of column density number ($N_e$) computed from the 
technique applied to the MHD simulation.}
\label{elecmaps}
\end{figure}
In Fig.\ref{elecmaps}a we show the actual column density computed from the MHD simulation
and in Fig.\ref{elecmaps}b we show the column density computed from 
synthesized METIS images assuming the plasma is all placed in the plane of sky.
The two column density distributions show
a high degree of similarity over the range of densities investigated here.
In general the magnitude of the column density is reproduced,
as well as its distribution and features.

\begin{figure}[!htcb]
\centering
\includegraphics[scale=0.20]{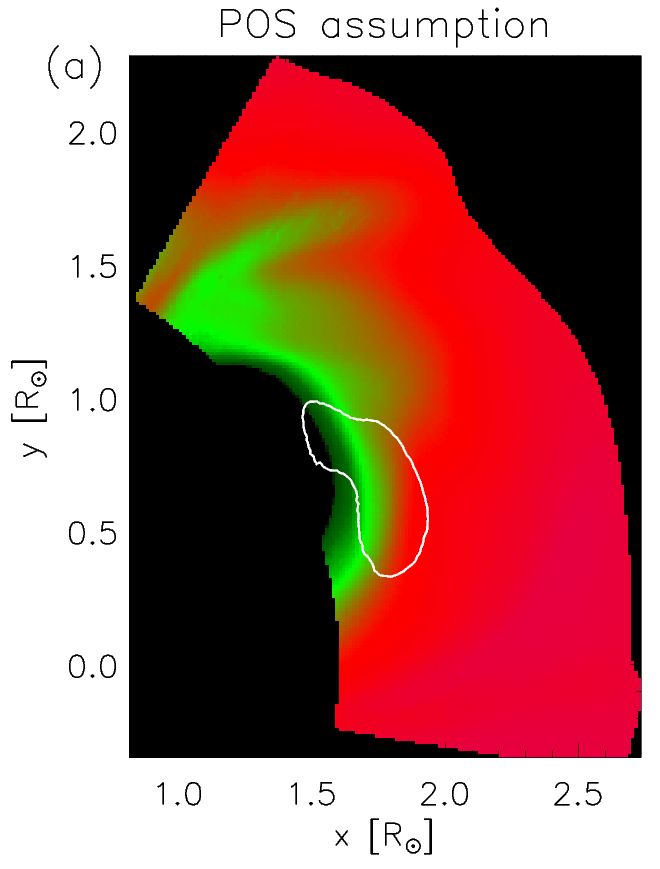} 
\includegraphics[scale=0.20]{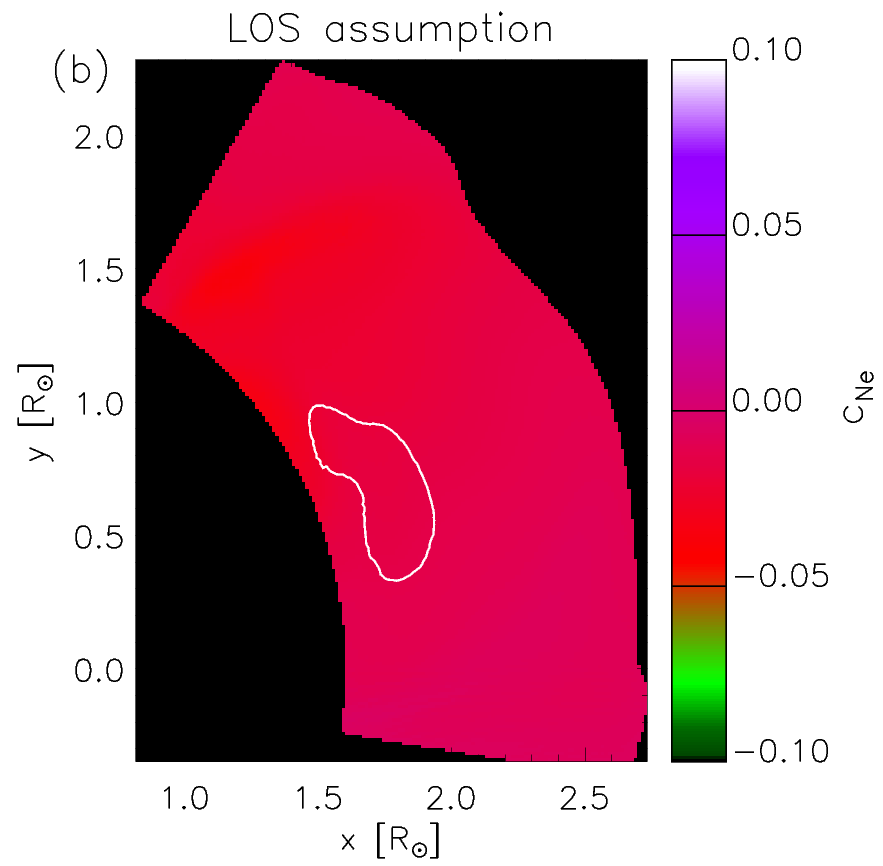} 
\caption{Maps of the contrast $c_{N_e}$ (Eq.\ref{contrastn}) between column density computed by synthesis of white-light images
and the actual column density from the MHD simulation under two different assumptions:
(a) all emitting plasma is located on the plane of sky;
(b) all emitting plasma is located at the distance from the plane of sky computed by the polarization ratio technique.
The white contour roughly identifies the flux rope position where the ratio
between the $y$ component of the magnetic field
integrated along the line of sight divided by the magnetic field intensity
is above the arbitrary positive threshold of 0.4.}
\label{elecmapserror}
\end{figure}
In order to illustrate the error
associated with this technique,
we map the contrast between the 
column density computed by the polarization ratio technique
and the actual column density computed from the MHD simulation in Fig.\ref{elecmapserror}.
The contrast $c_{N_e}$ is computed as
\begin{equation}
\label{contrastn}
c_{N_e}=\frac{N_e(WL)-N_e(MHD)}{N_e(MHD)}
,\end{equation}
where $N_e(WL)$ is the column density computed from the synthesis of white-light images
and $N_e(MHD)$ is the column density of the MHD simulation.
We consider the quantity $N_e(WL)$ under two assumptions:
i) the plasma is centred on the plane of sky (Fig.\ref{elecmapserror}a, ``POS assumption'')
and ii) the plasma is centred on the position along the LOS as computed by the polarization ratio technique (Fig.\ref{elecmapserror}b, ``LOS assumption'').
In both maps we also show the position of the flux rope (white contour).

The distribution of the error shows some interesting features.
We find that the POS assumption leads to an error of up to 10\% on the column density value.
In the region above the flux rope position the column density is generally underestimated by both techniques by about 5\%.
At the location of the flux rope
the underestimation is around 10\% for the POS assumption,
but reduces to $\sim3$\% in the LOS assumption.
The column density is only overestimated in a small location near the equator under the LOS assumption.
The underestimation near the flux rope is larger for the POS assumption (Fig.\ref{elecmapserror}a)
than for the LOS assumption (Fig.\ref{elecmapserror}b)  because the approximation of the latter
(a bulk of plasma at the position determined by the polarization ratio) is more precise than
the approximation of the former (a bulk of plasma at the plane of sky).
The same is true for all the locations that presented a dense feature in Fig.\ref{elecmaps},
as in general the relative error $c_{N_e}$ associated with the LOS assumption
is half of that associated with the POS assumption at any given location.
This interesting result suggests  an improved technique for the determination of column densities from white-light images of CMEs.
In particular, uncertainties can be minimized by assuming that in each pixel of the 2D coronagraphic image the emitting plasma
is not centred on the POS (as  is usually assumed in the literature),
but is centred on the actual location along the LOS as inferred from the polarization ratio technique.

\subsection{Time evolution}
By taking into account a series of snapshots from the MHD simulation,
processing them as in Sec.\ref{metisobservation} and
applying the same analysis described in Sec.\ref{foldedcentre} and Sec.\ref{densityestimate},
we can investigate the capabilities of the polarization technique 
in inferring the evolution of the CME.
In particular we can consider
its trajectory, velocity, expansion, and 
the mass involved as a function of time.

\begin{figure}[!htcb]
\centering
\includegraphics[scale=0.30]{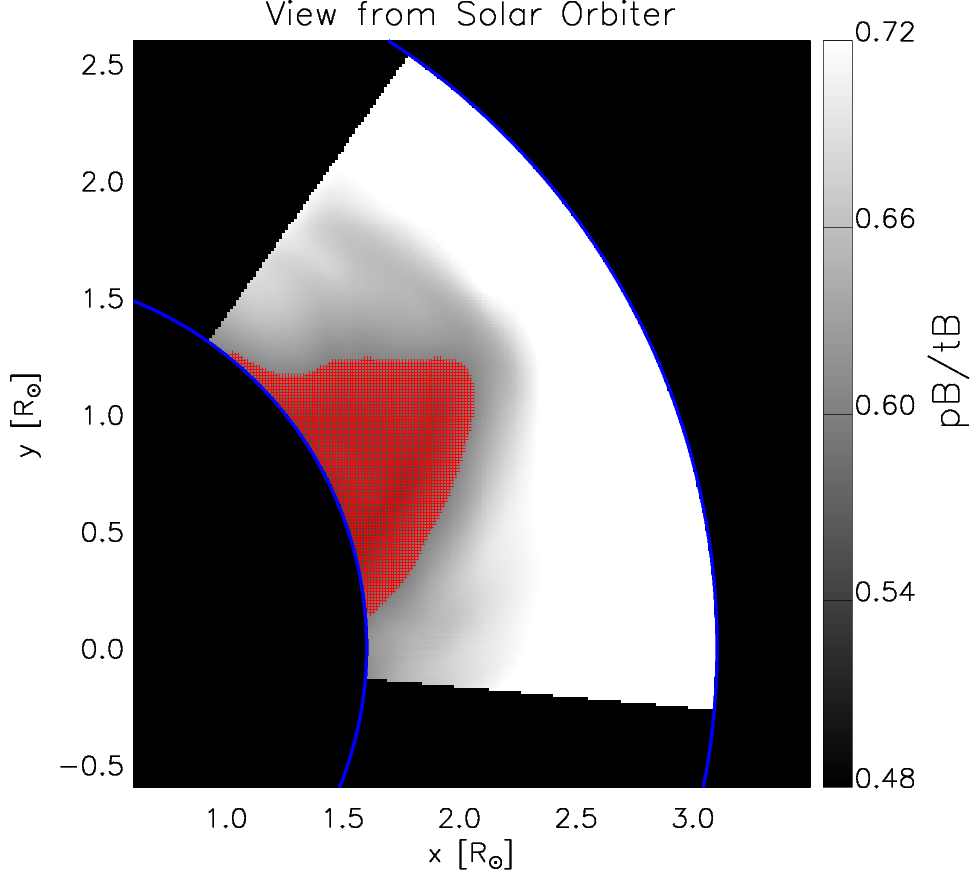} 
\caption{Map of polarization ratio ($pB/tB$);
 superimposed are the lines of sight that interject the CME according to our criteria ($pB/tB\le0.6$) at $t=127.6$ $min$.}
\label{trajectoryz10}
\end{figure}

In contrast to Sec.\ref{metisobservation}, we now want to apply
the analysis to lines of sight that cross the CME,
and to do so we need a way to identify them.
We decided this was the best choice because the polarization ratio technique performs more accurately
when applied to lines of sight that cross dense structures, which are the most relevant for our study.
The polarization ratio pB/tB of the white-light emission from a single structure 
is highest on the POS and decreases with distance  from the POS.
In our simulation the CME is composed of structures that do not lie on the POS.
We use this property to identify the CME in the LOS integrated images (panels in Fig. \ref{pol0060})
synthesized from the snapshots of the MHD simulation.
The CME cloud is identified as the region where the polarization ratio of the LOS integrated images
is less than a critical value of pB/tB $= 0.6$.
It can be shown that beyond a certain heliospheric distance,
the ratio pB/tB of the emission decreases while a 
structure moves radially (Fig.6 in \cite{BemporadPagano2015}).
Therefore the use of a constant critical upper value for pB/tB is a reasonable technique for   identifying
structures that are mostly radially propagating, like CMEs.
A map of polarization ratio with the considered lines of sight at $t=127.6$ $min$ is shown in Fig.\ref{trajectoryz10}.
From this point of view,
the CME appears as an expanding cloud wider near the solar surface with
a cusp that represents the farthest distance reached by the CME.
This method is applied ad hoc for this case in order to describe a smooth CME evolution
and remove spurious features from the analysis.
It does not aim at being a general method for CME detection in coronagraph images.
Different approaches are possible, and more complex detection techniques exist
\citep[e.g.][]{Bonte2011}.
From the polarization ratio we know that the CME enters
the METIS field of view after only $\sim80$ $min$
and stays mostly inside the field of view for an additional $60$ $min$.
Consequently we consider in our analysis only snapshots within this time range.

\begin{figure}[!htcb]
\centering
\includegraphics[scale=0.28]{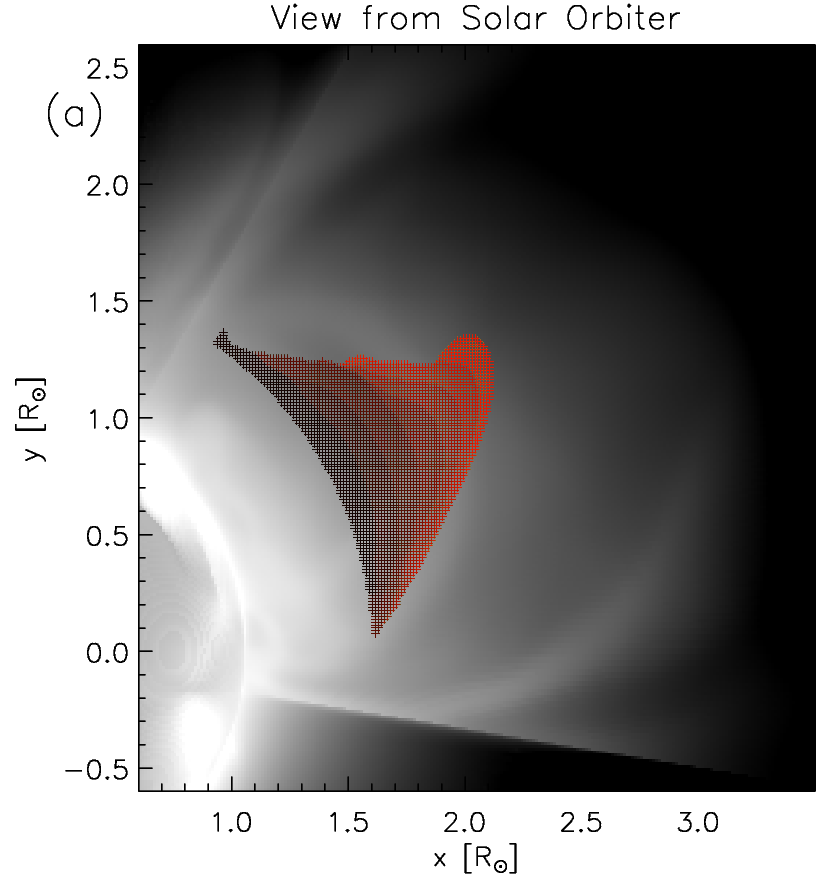} 
\includegraphics[scale=0.28]{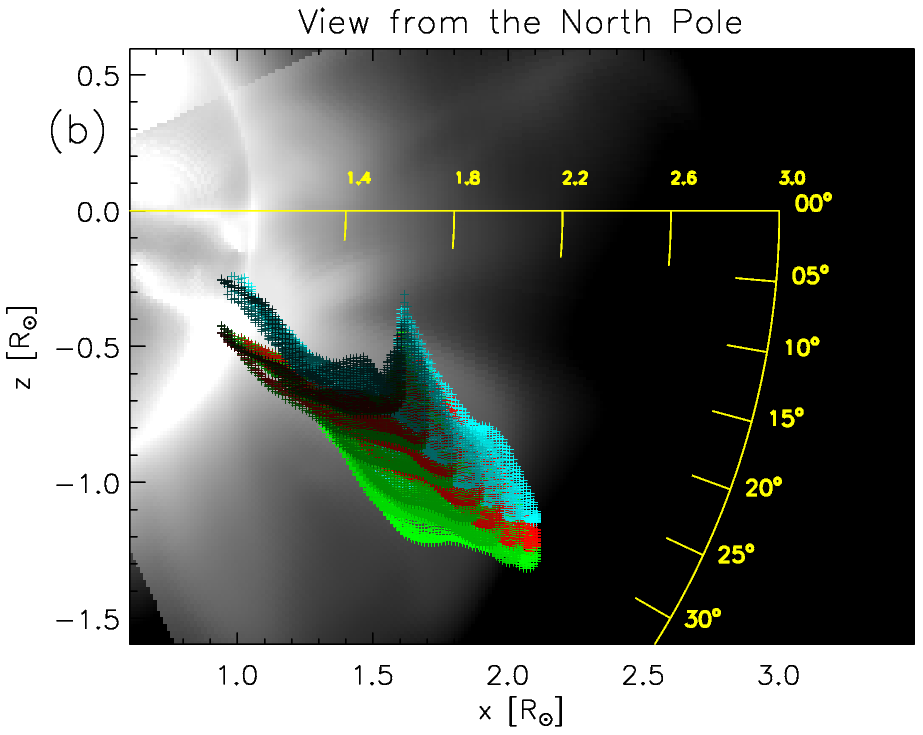} 
\includegraphics[scale=0.28]{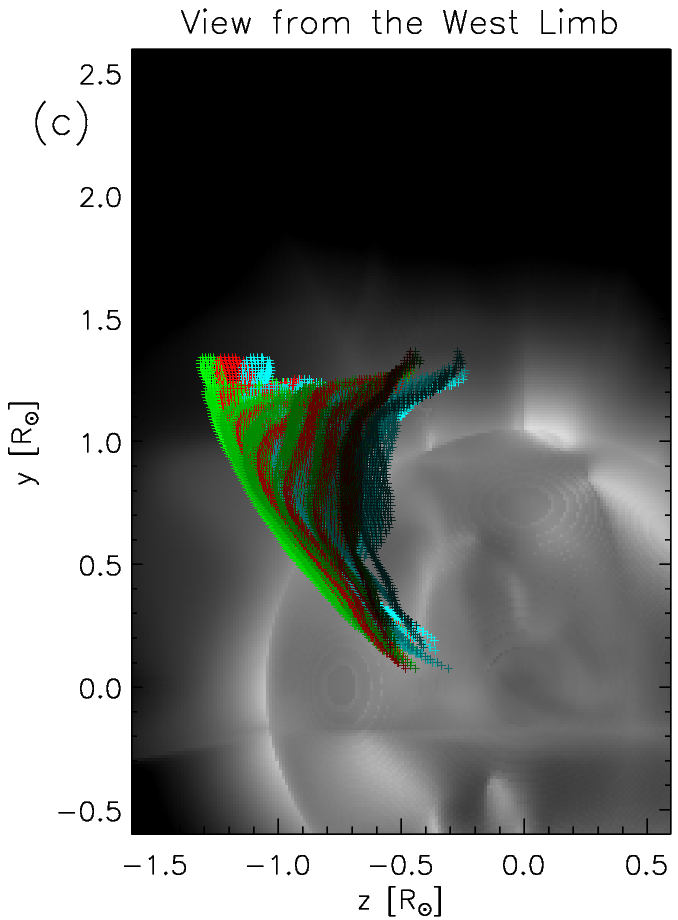} 
\caption{Maps of $Log_{10}$ of column density
(range between $10^{-15}$ $g/cm^2$ and $10^{-12}$ $g/cm^2$)
seen from three different points of view;
  superimposed  are the points of the clouds that belong to the CME.
Points obtained from the polarization ratio technique are red,
points of the folded centre of mass are green,
and points of the centre of mass are light blue.
(a) Point of view parallel to direction $z$ and $(x,y)$ coordinates of clouds of points;
(b) point of view parallel to direction $y$ and $(x,z)$ coordinates of clouds of points;
(c) point of view parallel to direction $x$ and $(z,y)$ coordinates of clouds of points.
A movie of these panels is available online.
}
\label{trajectoryy}
\end{figure}
Figure \ref{trajectoryy} shows the CME cloud of points at different times
from three different points of view (POS, north pole, west limb).
We use a different colour intensity for each cloud to represent different snapshots (from dark to bright),
so that the time evolution can be perceived.
Figure \ref{trajectoryy}a shows that our CME clouds are composed of a compact front
when the CME just enters the METIS field of view and it evolves into a less compact
front with a triangular shape.
The time evolution of the three clouds generally confirm the analysis carried out in Sec.\ref{result}.
From the Sun's north pole view (Fig.\ref{trajectoryy}b) we see that the CME keeps a narrow shape along the $z$-direction and it propagates at farther 
radial distances with a coherent motion.
Figure \ref{trajectoryy}b also shows that the clouds of points from the folded centre of mass and from the polarization ratio technique mostly coincide
with a small offset that displaces the folded centre of mass slightly farther from the POS than the polarization technique.
As we found in Fig.\ref{clouddiff}, the lines of sight that cross the flux rope
show an inverted behaviour with regard to the other lines of sight.
Namely that the folded centre of mass is farther from the POS than the position obtained from the polarization technique.
This is the reason why the relative position between the red and green clouds is switched between Fig.\ref{trajectoryy} and Fig.\ref{cloudszyx}.
This happens as the flux rope structure is significantly denser than the background corona and it lies at an angle larger than $20^{\circ}$
from the POS than is shown in Fig.\ref{cloudszyx}d. 

At the same time, both clouds (green and red) are displaced significantly farther from the POS than the true centre of mass (light blue).
Figure\ref{trajectoryy}c shows again the offset along the $z$ direction and also that the three clouds do not present any significant discrepancy
along $y$. 
It also reveals the rough bow shape of the propagation front of the CME.

\begin{figure}[!htcb]
\centering
\includegraphics[scale=0.30]{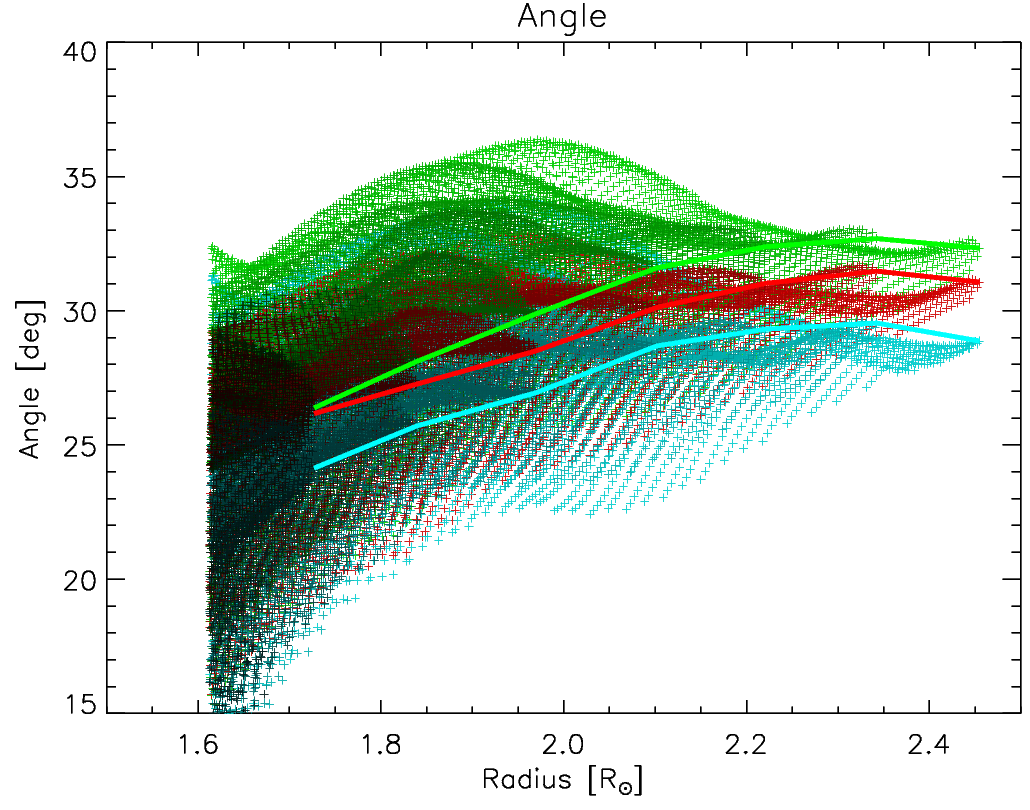} 
\caption{Angle from the POS as a function of the projected heliospheric distance for the clouds of points
as in Fig.\ref{trajectoryy}.}
\label{trajectoryangle}
\end{figure}
Similarly, we plot in Fig.\ref{trajectoryangle} the angle from the POS as a function of the heliocentric distance projected onto the POS
in order to show the progression of the CME and the accuracy of the polarization technique in catching its trajectory.
In order to trace a rough trajectory of the CME,
we represent each snapshot by the LOS with the maximum projected radial distance
and its angle with the POS (red, green, and light blue lines in Fig.\ref{trajectoryangle}).
We find that all the clouds follow the same pattern, where the CME is deflected by $\sim 5^{\circ}$.
The trajectory of the cloud of the centre of mass (light blue) starts at $\sim24^{\circ}$ (where the flux rope is placed)
and it ends at $\sim29^{\circ}$. 
The trajectories of the clouds of the polarization ratio technique and folded centre of mass
show an offset of about $2^{\circ}$ and $3^{\circ}$, respectively, from this position.
From this estimate of the trajectory, we can infer a CME velocity
of about $170$ $km/s$ and a velocity projected onto the POS of $130$ $km/s$.
This has to be compared with the velocity of the front of the CME as computed in Fig.\ref{centerfrposition},
where we found that the speed of the flux rope was of $157$ $km/s$.

\begin{figure}[!htcb]
\centering
\includegraphics[scale=0.25]{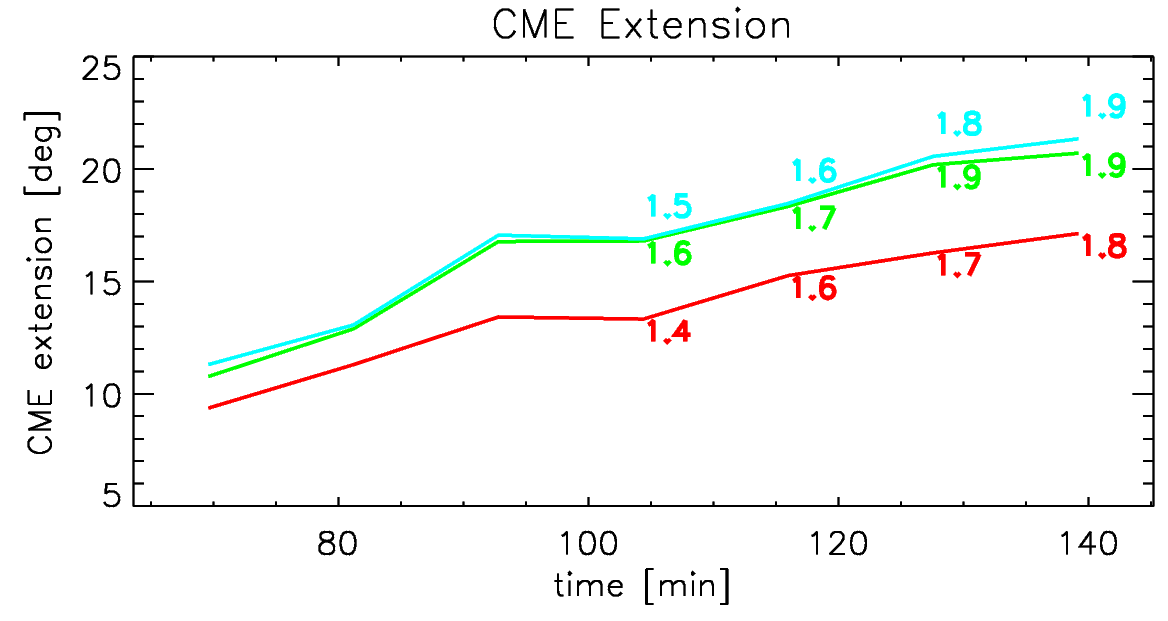} 
\caption{Angular extension of the CME as a function of time according to the different clouds.
Superimposed are the expansion factors at certain times.}
\label{expansion}
\end{figure}
Figure \ref{expansion} shows the angular extension of the CME for the three clouds as a function of time
where we also report the expansion factor at given times.
The expansion factor is simply the ratio between the extension of the cloud and its initial extension.
We find that the centre of mass cloud and the folded centre of mass cloud 
always show a similar extension,
while the polarization technique cloud is always narrower.
However, this does not significantly affect the measurement of the expansion,
which in the end differs by only $0.1$.
The folded centre of mass and centre of mass clouds
present an expansion about 5\% larger than the polarization technique cloud.

\begin{figure}[!htcb]
\centering
\includegraphics[scale=0.25]{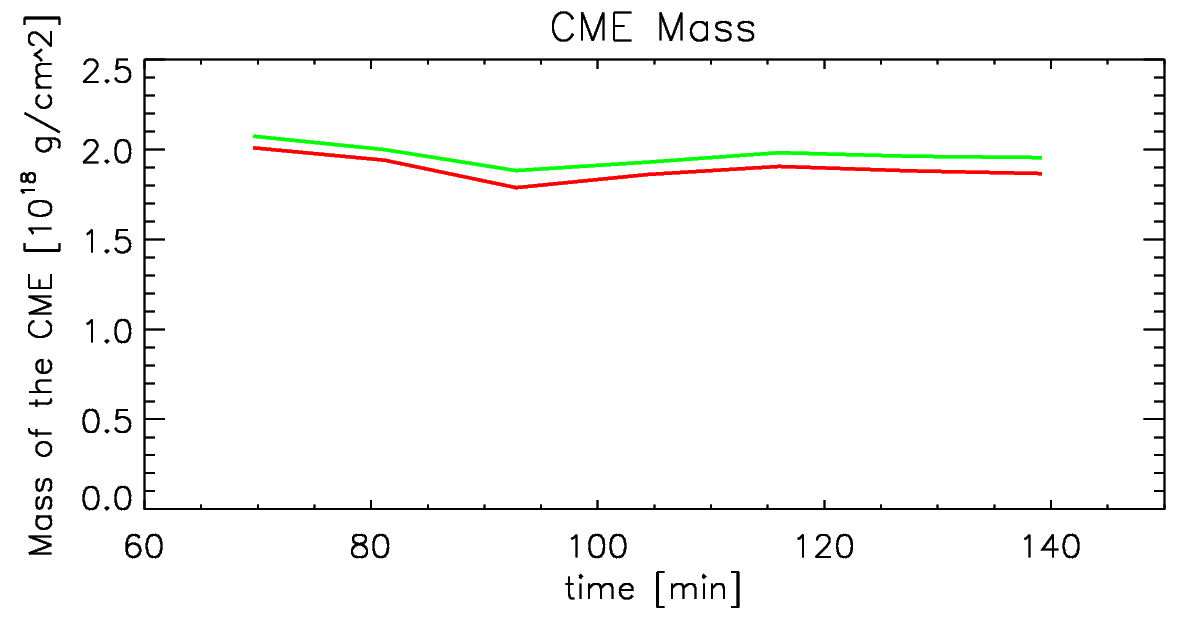} 
\caption{Average column density per line of sight as a function of time.}
\label{masscme}
\end{figure}
Finally, we can sum the column density measured by the polarization technique over the CME lines of sight
and compare it with the actual column density along the same lines of sight in the MHD simulation.
This will allow us to assess the mass variation of the CME as a function of time.
During its propagation, as the CME expands in the METIS FOV, it covers a wider area and
the total column density involved in the CME at the end of the simulation is 5 times greater
than the mass involved when it enters the METIS FOV.
Since the mass measurement is greatly affected by the number of lines of sight that we consider belong to the CME,
Fig.\ref{masscme} shows the total column density of the entire CME
as function of time
where it is divided by the number of CME points at each time.
It shows the value computed
from the polarization technique under the LOS assumption (green line)
and from the MHD simulation (red line).
The averaged column density steadily decreases and by the end it is $\sim92\%$ less than the value found at the beginning.
However, the discrepancy between the two measurements is small where its maximum value corresponds
to only $\sim3\%$ of the average column density.
This proves once more that the polarization technique is a reliable technique for  measuring the column density along a line of sight.

\section{Discussion and conclusions}
\label{discussion}

In this work we have investigated the reliability of results of the 3D structure of CMEs as inferred
from the polarization ratio technique.
The technique was applied to white-light images of total (tB) and polarized (pB) brightness
synthesized from a MHD simulation of a flux rope eruption.
For synthesis purposes we used the METIS coronagraph field of view.
By comparing the real plasma density distribution with the 
results from the polarization ratio technique,
we analyse how the position, direction of propagation of the CME,
and the column density of plasma are reconstructed.

\subsection{Observations of CMEs by METIS/Solar Orbiter}
The synthesis of the white-light images as seen by METIS
from the MHD simulation allows us to show possible future observations of
CMEs when the Solar Orbiter spacecraft will reach  perihelion at $0.29$ AU.
At this heliocentric distance the projected field of view of METIS will extend down to
$1.8$ $R_{\odot}$ from the solar surface.
This will observe an altitude where in general ejected flux ropes have not yet merged in the solar wind.
Hence, METIS will allow us to carry out coronagraphic observations
of fully formed ejected flux ropes in both visible and UV ranges.
Furthermore, both METIS (imaging the solar corona) and EUI (imaging the solar disk)
will observe with the Ly-$\alpha$ intensity.
Solar Orbiter will therefore provide a unique opportunity to follow the evolution of flux ropes  from their formation
to ejection out to about $3$ $R_{\odot}$
where the merging of the flux rope with the solar wind eventually occurs.
At the same time, the METIS field of view will be large enough
to catch in the same image the CME front and other related phenomena such as CME-driven shocks and side interactions - reconnection with nearby coronal structures.
With the spatial resolution of the instrument (20 arcsec/pixel)
enhanced by the proximity of the satellite to the Sun,
it will be possible to describe with unprecedented
clarity the density distribution of the three components of CMEs.
In addition we will be able to derive information on the CME plasma temperature by combining
for example white-light and UV images acquired at the same time.
In particular, this latter point will be the subject of future work where
synthetic white-light and Ly-$\alpha$ METIS images of CMEs will be analysed
in order to enhance our diagnostic capacity and 
to develop techniques that will
be available for when actual data is available.
In the work presented here, we use a MHD simulation and the synthesis of visible-light images to show preliminary aspects of how METIS will observe CMEs when in operation.
Additionally, it is worth considering that Solar Orbiter will travel on an orbit not co-planar with the Earth's orbit.
Although we have carried out our analysis from a point of view on the Earth's orbit, 
it is possible to imagine that the polarization technique can be used
to study the 3D structure of CMEs from points of view
off of Earth's orbit plane.

\subsection{3D Reconstruction of a CME with polarization ratio}
\label{disc3drecon}
Overall, the results from the polarization ratio technique
 successfully  infer the 3D CME structure
once we have a clear understanding of the
interpretation of results and the errors associated.
When applied to the present MHD simulation, the polarization ratio technique
is able to infer the location of the flux rope and other density structures
with sufficient precision to describe the general structure of the CME.
In particular we confirm here, as already discussed in \citet{BemporadPagano2015},
that the polarization ratio technique provides a good estimate for
the location along the line of sight (LOS) of what we call the folded centre of mass,
namely,  the centre of mass of the density distribution obtained by summing
the distribution behind the plane of the sky (POS) to the one in front of it.
In particular, the technique measures the position of the folded centre of mass within an error of $\sim0.03$ $R_{\odot}$.
This is a relatively small error considering that the extension of a CME can be over
$1$ $R_{\odot}$ in the radial direction.
Moreover, considering a CME velocity on the order of $500$ $km/s$,
a distance of $\sim0.03$ $R_{\odot}$ is covered in approximately $40$ seconds,
while the CME event lasts up to a few hours.
Thus, from this point of view, the use of the polarization technique
would lead to a minimal error if applied
to the determination of large-scale properties such 
as the injection time of the CME 
into the solar wind or the arrival time at the Earth's magnetosphere.

\begin{figure}[!htcb]
\centering
\includegraphics[scale=0.30]{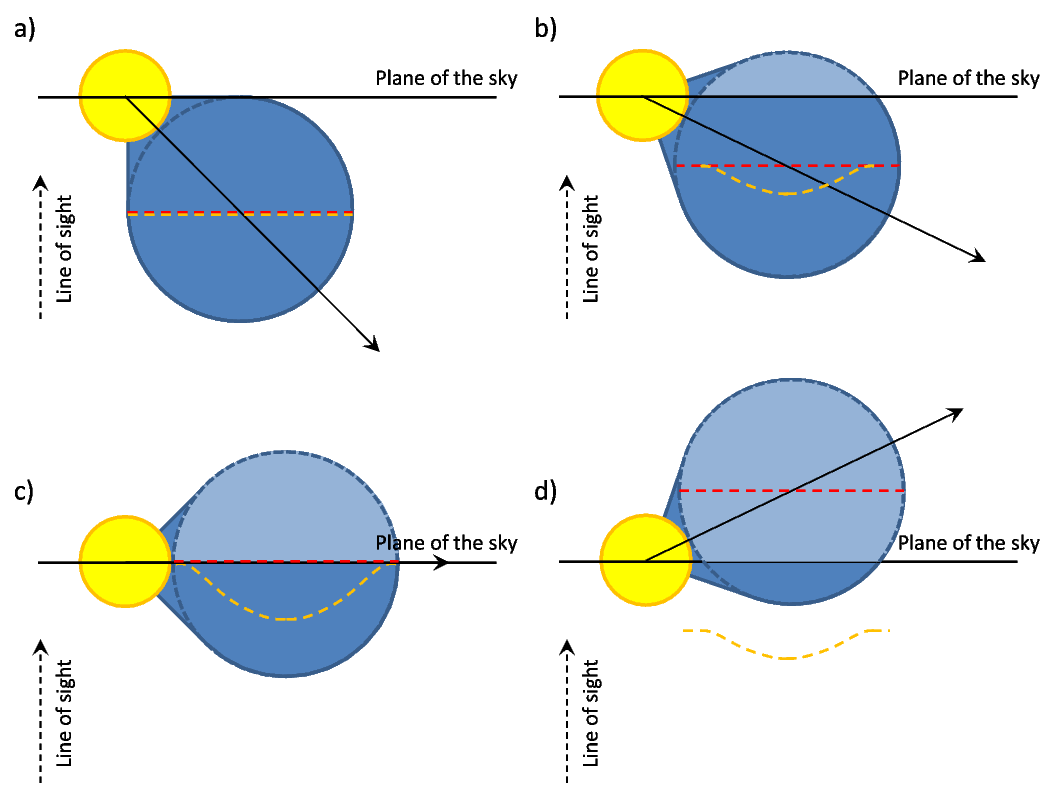} 
\caption{Idealized cartoon showing the location of the real centre of mass (red dashed line)
and of the folded density centre of mass (orange dashed line) for different orientations of the CME with respect to the POS (see text).}
\label{cartoon}
\end{figure}

Nevertheless, it should be noted that such a precision is realistic only when
the folded centre of mass and the centre of mass are placed approximately at the same location.
This only occurs when the CME travels and expands without crossing the POS.
If, however, the CME bubble crosses the POS the errors introduced by the polarization ratio technique
may be up to $\sim10$ times larger.
This is shown  in Fig.\ref{cartoon}
where the CME bubble (assumed  to be spherical and with homogeneous density)
lies entirely out of the POS (top left)
the location of the real centre of mass (red dashed line) is coincident
with the centre of mass of the folded density distribution and with 
the one inferred from the polarization ratio technique (orange dashed line).
Nevertheless, when the CME bubble partially crosses the POS (top right)
a disagreement between the two distributions starts to appear.
The disagreement is larger at the centre of the CME
where the observed tB and pB intensities come from a longer integration along the LOS.
The uncertainty maximizes when the CME expands along the POS (bottom left);
due to the symmetry in Thomson scattering geometry,
it is not possible to distinguish the case when the CME is expanding in front of (top right) or behind (bottom right) the POS.
It should be noted here that the estimation of the errors associated with the polarization ratio technique
that we find here, using a realistic MHD simulation are in agreement with those derived in \citet{BemporadPagano2015}.
This is an important cross-check that validates our previous analysis.
\citet{Dai2014} have also concluded that the polarization ratio technique is affected
by an ambiguity in identifying the location of the source of emission and have proposed a method to overcome this.
Our work does not tackle the problem from this general point of view, but we focus on the errors that persist
in the measurements after the ambiguity is removed.
This is especially true for the case they call the implicit ambiguity.

\subsection{Estimate of the CME column density}
We find that the measurement of column density
obtained from white-light images
generally matches the actual column density of the MHD simulation to within $10\%$.
The measurements are carried out by assuming that the entire column of plasma
is located at one location along the LOS,
either on the POS (as usually done in data analysis),
or at the position of the centre of mass as inferred from the polarization ratio technique.
Our study shows that when we assume that all the plasma is located 
at the position inferred from the polarization ratio we get a more precise measurement of the column density.
This is particularly true near the flux rope location which is
the more interesting region for estimating the mass of the CME.
In our study the error reduces by a factor of 3 when the LOS assumption is used.
Hence, this can be considered  an improved technique for the determination of column densities from white-light images of CMEs.
The uncertainties can be minimized by assuming that in each pixel of the 2D coronagraphic image the emitting plasma
is centred on the position along the LOS inferred from the polarization ratio technique.

\subsection{Time evolution}
Finally, when we apply the comparison to a set of snapshots from the MHD simulation, 
we first need to identify the lines of sight that cross the CME
in order to apply the polarization technique to a sensible region of the FOV
and to extract information such as its trajectory or expansion rate. 
The polarization technique can show several dynamic features, 
but not all of them belong to a coherent and structured ejection like a CME.
Provided that the relevant lines of sight are identified,
the polarization ratio technique is able to reproduce the trajectory 
of the CME and its extension with the same spatial approximation that we have found when focusing on a single frame.
In our study, the offset between the polarization technique cloud and the centre of mass cloud
leads to an error in the CME trajectory of $5^{\circ}$ in the longitudinal coordinate.
This is a significant discrepancy that could be reduced with more complex data analysis
that takes into account the position on the solar disk where the CME originates.
When the initial offset is known, it is possible to infer the correct trajectory.
The error introduced on the speed of the CME is more difficult to infer because it is closely connected with the detection technique that determines the position and thus the velocity of the CME front.
On the other hand, the shape of the CME cloud and its evolution is captured by the polarization technique.
The polarization ratio technique enhances a more accurate measurement of mass involved in a CME.
Especially by using the LOS assumption the technique returns
a column density accurate to within $2\%$ when a CME crosses the field of view.

\subsection{Application of the technique for space weather forecast}
As the polarization ratio technique is able to provide accurate measurements of
the mass involved in a CME (through the column density estimate)
and the position and velocity of the CME (through the centre of mass at different times),
we can imagine using this technique to detect the early stages of CMEs directed towards the Earth
and with the support 
of space forecasting models to infer the arrival time and the mass.
However, as we outlined in Sec.\ref{disc3drecon}, the polarization technique
introduces some errors in the localization of the CME that need to be reduced
so that the technique gives the best results.
The technique is much more accurate at localizing the folded centre of mass
than the real centre of mass.
Also, we have highlighted that this error increases the more the CME crosses the POS.

\begin{figure}[!htcb]
\centering
\includegraphics[scale=0.4]{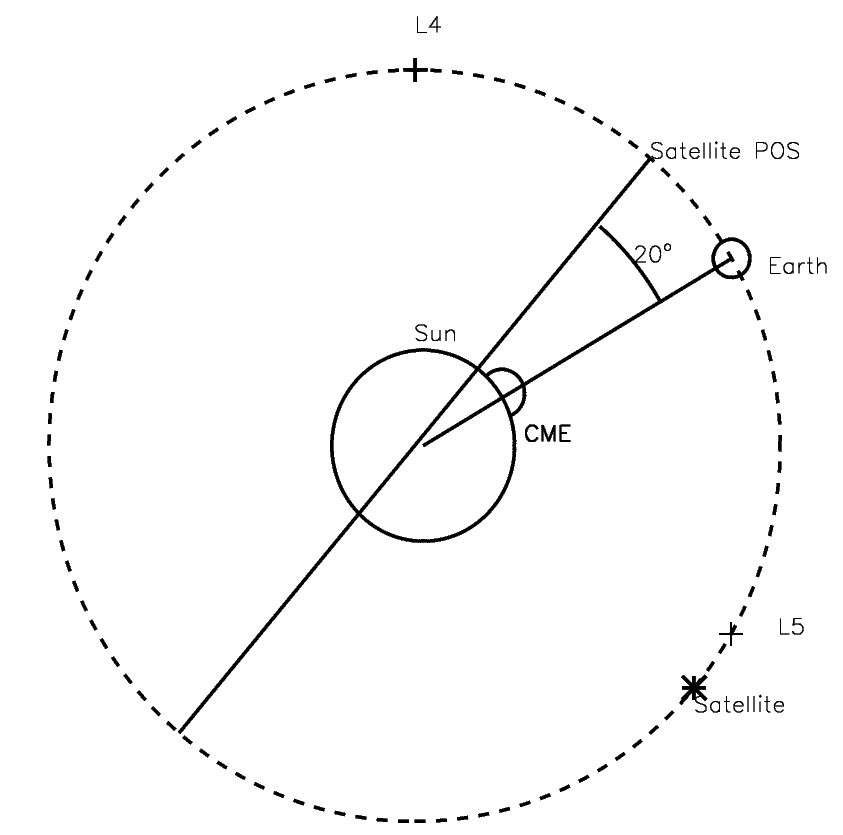} 
\caption{Sketch of the ideal configuration for a satellite to continuously monitor the initiation of CMEs
from the solar corona and to apply the polarization ratio technique
to compute the 3D structure and the speed of the ejection (not to scale).}
\label{satellite}
\end{figure}
As explained in \citet{BemporadPagano2015} and confirmed in this work,
the technique generally performs best when
the CME travels at about $20^{\circ}$ off of the plane of sky.
This peculiar position not only minimizes the error due to plasma crossing the plane of sky,
but also the error introduced by applying the technique at larger radial distance.
We therefore suggest that an ideal setup to timely and accurately detect CMEs
and measure their position, speed, and mass is to place a satellite 
that observes a large field of view in white light (both total and polarized brightness)
on an Earth orbit travelling $70^{\circ}$ ahead or behind the Earth, as in Fig.\ref{satellite}.
Within this orbit the direction of propagation of CMEs directed towards the Earth
makes an angle of $20^{\circ}$ with the POS.
This setup would allow us to continuously monitor the portion of the corona that
generates Earth direct CMEs and to promptly alert advanced space weather forecast systems
of the injection of a CME into interplanetary space with accurate estimation of its
mass, position, and speed.
It should be noted that the $L4$ and $L5$ Lagrangian points are
approximately located at a $60^{\circ}$ angle
in front of and behind the Earth and thus 
quite close to the ideal position for the 3D reconstruction of CMEs with the polarization ratio technique.
This is a significant and additional advantage to be considered for future space weather
monitoring missions that will be located in the $L4$ or $L5$ Lagrangian points.

\begin{acknowledgements}
We would like to thank the anonymous referee for the careful and constructive work that has improved the quality and the relevance of the manuscript.
We acknowledge the use of the open source (gitorious.org/amrvac) MPI-AMRVAC software, relying on coding efforts from C. Xia, O. Porth, R. Keppens.
DHM would like to thank STFC, the Leverhulme Trust and the European Commission's Seventh Framework Programme
(FP7/2007-2013)  for their financial support.
PP would like to thank STFC, the Leverhulme Trust, the European Commission's Seventh Framework Programme
(FP7/2007-2013) under grant agreement SWIFF (project 263340, www.swiff.eu).
The computational work for this paper was carried out on the joint STFC and SFC (SRIF) funded cluster at the University of St Andrews (Scotland, UK).
\end{acknowledgements}

\bibliographystyle{aa}
\bibliography{ref}

\end{document}